\documentstyle[12pt]{article}
\input epsf.sty

\renewenvironment{thebibliography}[1]
 {\begin{list}{\arabic{enumi}.}
    {\usecounter{enumi} \setlength{\parsep}{0pt}
     \setlength{\itemsep}{3pt} \settowidth{\labelwidth}{#1.}
     \sloppy
    }}{\end{list}}

\input epsf.sty
\input psfig.sty

\def\nn{\noindent}

\def\be{\begin{equation}}
\def\barr{\begin{array}}
\def\earr{\end{array}}

\def\gsim{\:\raisebox{-0.5ex}{$\stackrel{\textstyle>}{\sim}$}\:}

\def\etal{ {\it et al.} }

\def\ib#1,#2,#3{           {\it ibid.\/ }{\bf #1} (19#2) #3}
\def\ap#1,#2,#3{           {\it Ann. Phys. (NY)\/ }{\bf #1} (19#2) #3}
\def\ijmp#1,#2,#3{         {\it Int. J. Mod. Phys.\/ } {\bf A#1} (19#2) #3}
\def\mpl#1,#2,#3 {          {\it Mod. Phys. Lett.\/ } {\bf A#1} (19#2) #3}
\def\np#1,#2,#3{           {\it Nucl. Phys.\/ }{\bf B#1} (19#2) #3}
\def\npps#1,#2,#3{         {\it Nucl. Phys. B (Proc. Suppl.)\/ }{\bf B#1}
                             (19#2) #3}
\def\plb#1,#2,#3{           {\it Phys. Lett.\/ }{\bf B#1} (19#2) #3}
\def\pr#1,#2,#3{           {\it Phys. Rev.\/ }{\bf D#1} (19#2) #3}
\def\prep#1,#2,#3{         {\it Phys. Rep.\/ }{\bf #1} (19#2) #3}
\def\prl#1,#2,#3{          {\it Phys. Rev. Lett.\/ }{\bf #1} (19#2) #3}
\def\pro#1,#2,#3{          {\it Prog. Theor. Phys.\/ }{\bf #1} (19#2) #3}
\def\rmp#1,#2,#3{          {\it Rev. Mod. Phys.\/ }{\bf #1} (19#2) #3}
\def\sp#1,#2,#3{           {\it Sov. Phys.-Usp.\/ }{\bf #1} (19#2) #3}
\def\zp#1,#2,#3{           {\it Zeit. f\"ur Physik\/ }{\bf #1} (19#2) #3}

\newcommand{\cm}{center of mass}

\newcommand{\sw}{\mbox{$\sin\theta_w$}}
\newcommand{\cw}{\mbox{$\cos\theta_w$}}
\newcommand{\swt}{\mbox{$\sin^2\theta_w$}}

\newcommand{\pe}{\mbox{$e^+e^-$}}

\newcommand{\ep}{\mbox{$e^-\gamma$}}

\newcommand{\dil}{bilepton}

\begin{document}
\thispagestyle{empty}

\newlength{\captsize} \let\captsize=\small % use \let\normalsize=\captsize
%\newlength{\captwidth}                     % just before \caption{  ...

\font\fortssbx=cmssbx10 scaled \magstep2
\hbox to \hsize{
%\special{psfile=iu_logo.ps
%hscale=15 vscale=15
%hoffset=-12 voffset=-2}
%\hskip.5in
\raise.1in\hbox{\fortssbx Indiana University - Bloomington}
\hfill\vbox{\hbox{\bf IUHET-348}
            \hbox{November 1996}}
}

\vspace*{.5in}

\begin{center}
{\large\bf
Discovering New Particles
at 
Colliders
%\footnotemark
}\\[.1in]
\small
M.S.~Berger$^1$ and W.~Merritt$^2$
\\[.1in]
\small\it
$^1$Physics Department, Indiana University, Bloomington, IN 47405,
USA\\
\small\it
$^2$Fermi National Accelerator Laboratory, 
Batavia, IL 60510, USA\\
\end{center}

\vspace{.5in}

\begin{abstract}
We summarize the activities of the New Particles Subgroup at the 1996 Snowmass 
Workshop. We present the expectations for discovery or exclusion of leptoquarks
at hadron and lepton colliders in the pair production and single production
modes. The indirect detection of a scalar lepton quark at polarized $e^+e^-$ 
and $\mu^+\mu^-$ colliders is discussed. The discovery prospects 
for particles with two units of lepton
number is discussed. We summarize the analysis of the single production of 
neutral heavy leptons at lepton colliders.
\end{abstract}

\vspace{1.5in}

\noindent 
To appear in the {\it Proceedings of the 1996 DPF/DPB Summer Study on New 
Directions for High Energy Physics-Snowmass96}, Snowmass, CO, 25 June-12 July,
1996.

\newpage

\section{Introduction}

With the recent discovery of a new particle at the Tevatron, one of the last
holes in the Standard Model has been filled. Only the elusive Higgs boson 
remains undetected, and the next new particle 
discovery may usher in a revolution in our understanding.
The area covered by the subgroup was all new particles which are fundamental
and not covered by other groups at Snowmass. This includes fermions with
exotic (non-Standard Model) quantum numbers, sequential fermions (i.e. a fourth
generation), and leptoquarks and diquarks. Other new particles covered
by other working groups such as the Higgs boson and supersymmetric particles
were omitted from our studies. New non-fundamental (composite) particles 
such as excited fermions and technipions were
relegated to the New Interactions subgroup\cite{newint}.
A recent review of new particles and interactions can be found in 
Ref.~\cite{review}.

In this report we summarize the individual contributions to the New Particles
subgroup. For more details the individual contributions should be consulted.

\section{Leptoquarks}

Theories attempting to unify the leptons and quarks in some common framework
often contain new states that couple to lepton-quark pairs, and hence are 
called leptoquarks\cite{guts}. Necessarily leptoquarks are color triplets,
carry both baryon number and lepton number, and can be either spin-0
(scalar) or spin-1 (vector) particles.
Perhaps the most well-known examples of leptoquarks appear as gauge
bosons of grand unified theories\cite{gg}. To prevent rapid proton decay they
must be very heavy and unobservable, or their couplings must be 
constrained by symmetries. Nonetheless, much work has been devoted 
to signals for the detection of leptoquarks at present and future 
colliders\cite{pp,ep,hr,ee,egam,gamgam}.
Searches have already been performed at
LEP{\cite {lep}}, HERA{\cite {hera}}, and the Tevatron{\cite {tev}}.
One potentially attractive source of light leptoquarks is in $E_6$ models 
where the scalar leptoquark can arise as the supersymmetric partner to the 
color-triplet quark that naturally resides in the fundamental representation
{\bf 27}. A recent review of the physics signals for leptoquarks can be found
in Ref.~\cite{review}.

Leptoquarks can be sought by looking for indirect effects in low energy 
processes\cite{sacha}. Light leptoquarks (less than a several hundred GeV)
must also satisfy strong constraints from
flavor changing neutral current processes, so that leptoquarks must couple
to a single generation of quarks and leptons. 
The most convincing evidence for leptoquarks would come from their direct
production and detection at colliders. 
For the heavy leptoquarks that might be detected at the 
multi-TeV machines, the constraints
from low energy processes do not
necessarily require this, since the leptoquark's virtual contributions are 
suppressed by its large mass.

Various methods have been proposed to search for leptoquarks.
At lepton colliders ($e^+e^-$ and $\mu^+\mu^-$) colliders, leptoquarks can
be produced in pairs via $s$-channel $\gamma $ and $Z$ exchange, and by 
$t$-channel exchange of a quark. The coupling of a leptoquark is not 
constained by the usual gauge symmetries (it is a Yukawa coupling), so there
is some model dependence that 
necessarily enters in some cross section calculations. The production
cross section depends sensitively on the 
leptoquark couplings so that the constraints depend on its quantum numbers.

Leptoquarks decay into a lepton and a quark, giving quite distinctive 
signals. The signatures for leptoquark pair production are therefore:
(1) two charged leptons and two hadronic jets, 
(2) one charged lepton, two hadronic jets and missing energy (neutrino),
and (3) two hadronic jets and missing energy.
For relatively light leptoquarks, the constraints from flavor-changing 
neutral currents generally constrain the leptoquark couplings to be within
a single generation so that the leptons in the final state will be in the 
same family. For heavier leptoquarks ($M_{LQ}>1$~TeV) this is not 
necessarily the case and more exotic final states are possible.

Single production of leptoquarks is also possible. The cross sections for 
these processes generally depend on the unknown Yukawa coupling. The 
advantage in this case is that one can obtain a higher reach in leptoquark
mass since kinematically one only needs center-of-mass energy to make one
heavy particle.

Finally, one can look for virtual effects of leptoquarks (zero-production
of leptoquarks). In this case one can exclude leptoquarks in excess of the 
colliders center-of-mass energy by looking for deviations from the 
Standard Model predictions for cross sections and asymmetries.

\section{Pair Production of Leptoquarks in Hadron SuperColliders}

Rizzo\cite{rizzoproc} examined the search reach for both scalar and vector 
leptoquarks at future hadron supercolliders. The colliders considered
are the $\sqrt s$=60 (LSGNA) and 
200 (PIPETRON) TeV machines, operating in either a $pp$ or $p\overline{p}$ 
mode. At these energies and the anticipated luminosities
leptoquarks even above a TeV are accessible.

The dominant production for leptoquarks at a hadron collider is expected to 
be pair production, which proceeds 
through QCD interactions (in either $gg$ or $q\bar q$ collisions) and depends 
only on the leptoquark spin 
and the fact that it is a color triplet field\cite{scalar}. 

For vector leptoquarks ($V$), one can assume that they are the gauge bosons
of an extended gauge group. Then the $gVV$ and $ggVV$ couplings are fixed by 
extended gauge invariance.  
The Feynman rules needed for calculating the production cross section 
can then be derived from the following Lagrangian\cite {vlq}
\begin{equation}
{\cal L}_V =-{1\over 2} F^\dagger_{\mu\nu}F^{\mu\nu}+M_V^2V^\dagger_\mu V^\mu
-ig_sV^\dagger_\mu G^{\mu\nu}V_\nu  \,.
\end{equation}
Here, $G_{\mu\nu}$ is the usual gluon field strength tensor,
$V_\mu$ is the vector leptoquark field and $F_{\mu\nu}=D_\mu V_\nu-D_\nu
V_\mu$, where $D_\mu=\partial_\mu+ig_sT^a G^a_\mu$ is the gauge
covariant derivative (with respect to $SU(3)$ color), $G^a_\mu$ is the
gluon field and the $SU(3)$ generator $T^a$ is taken in the triplet
representation.  One can be more general than this i.e. not necessarily 
assuming that the leptoquark is a fundamental gauge boson. Then one can 
introduce an undetermined parameter $\kappa $ in the last term that acts as
an anomalous chromomagnetic moment; see Ref.~\cite{rizzoproc} for details.

\vspace*{-0.5cm}
\nn
\begin{figure}[htbp]
\centerline{
\psfig{figure=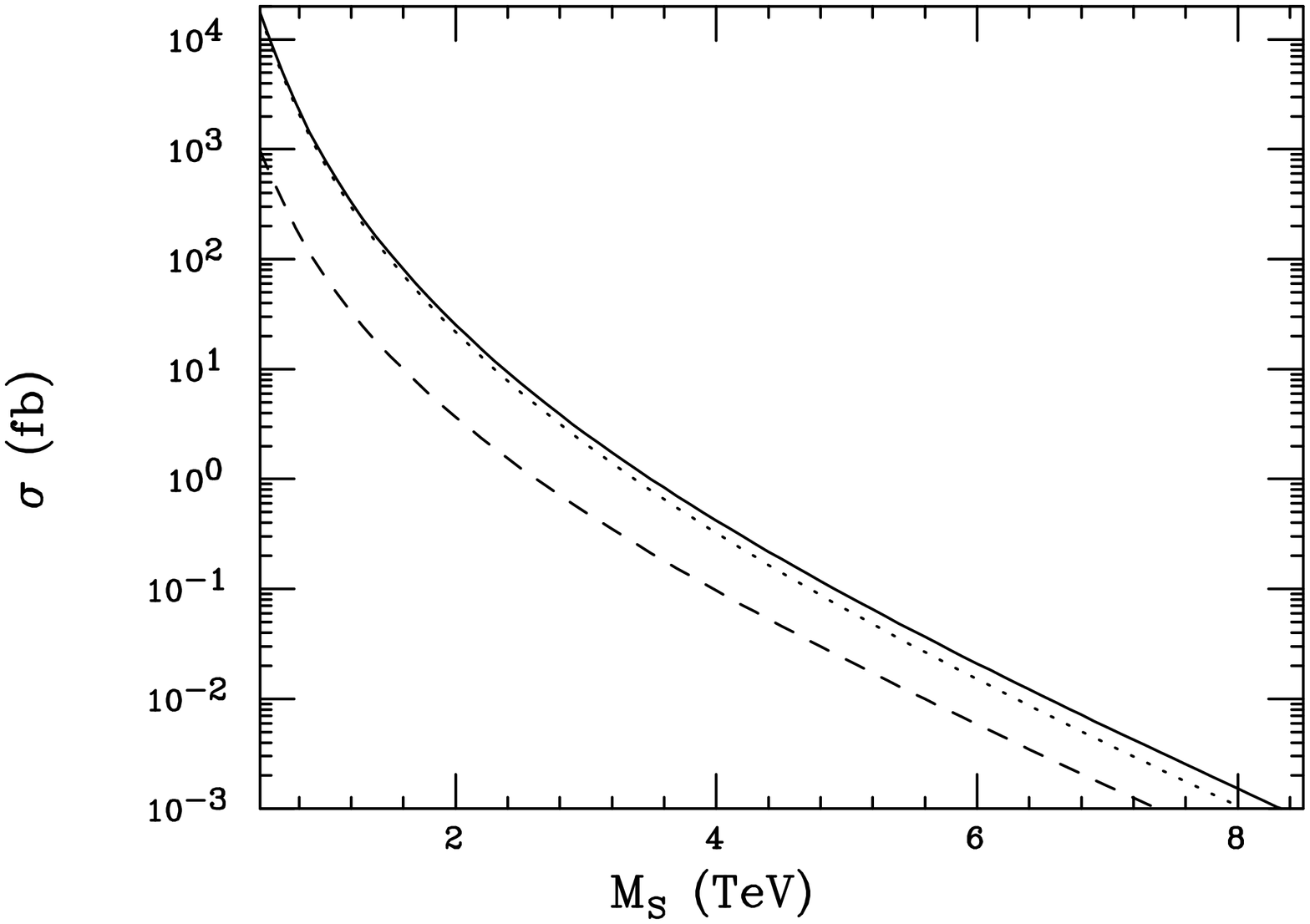,height=8.1cm,width=8.1cm,angle=-90}
\hspace*{-5mm}
\psfig{figure=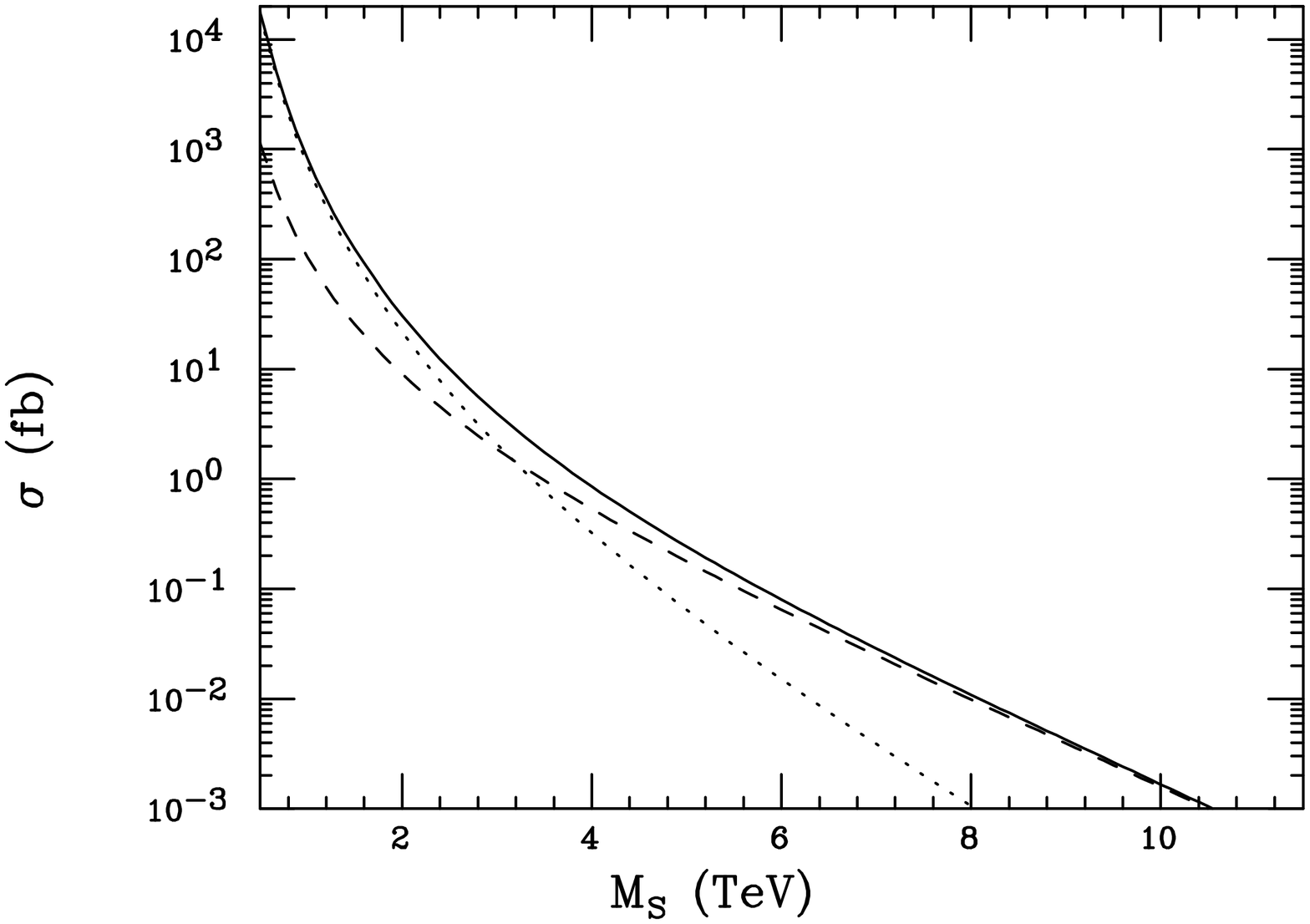,height=8.1cm,width=8.1cm,angle=-90}}
\vspace*{-0.6cm}
\caption{Scalar leptoquark pair production cross section as a function of mass 
at a 60 TeV $pp$(left) or $p\bar p$(right) LSGNA collider. The
dotted(dashed) curve corresponds to the $gg$($q\bar q$) production
subprocess whereas the solid curve is their sum. MRSA$'$ parton
densities are employed (from Ref.~{\protect\cite{rizzoproc}}).}
\label{s60}
\end{figure}
\vspace*{0.1mm}
\vspace*{-0.5cm}
\nn
\begin{figure}[htbp]
\centerline{
\psfig{figure=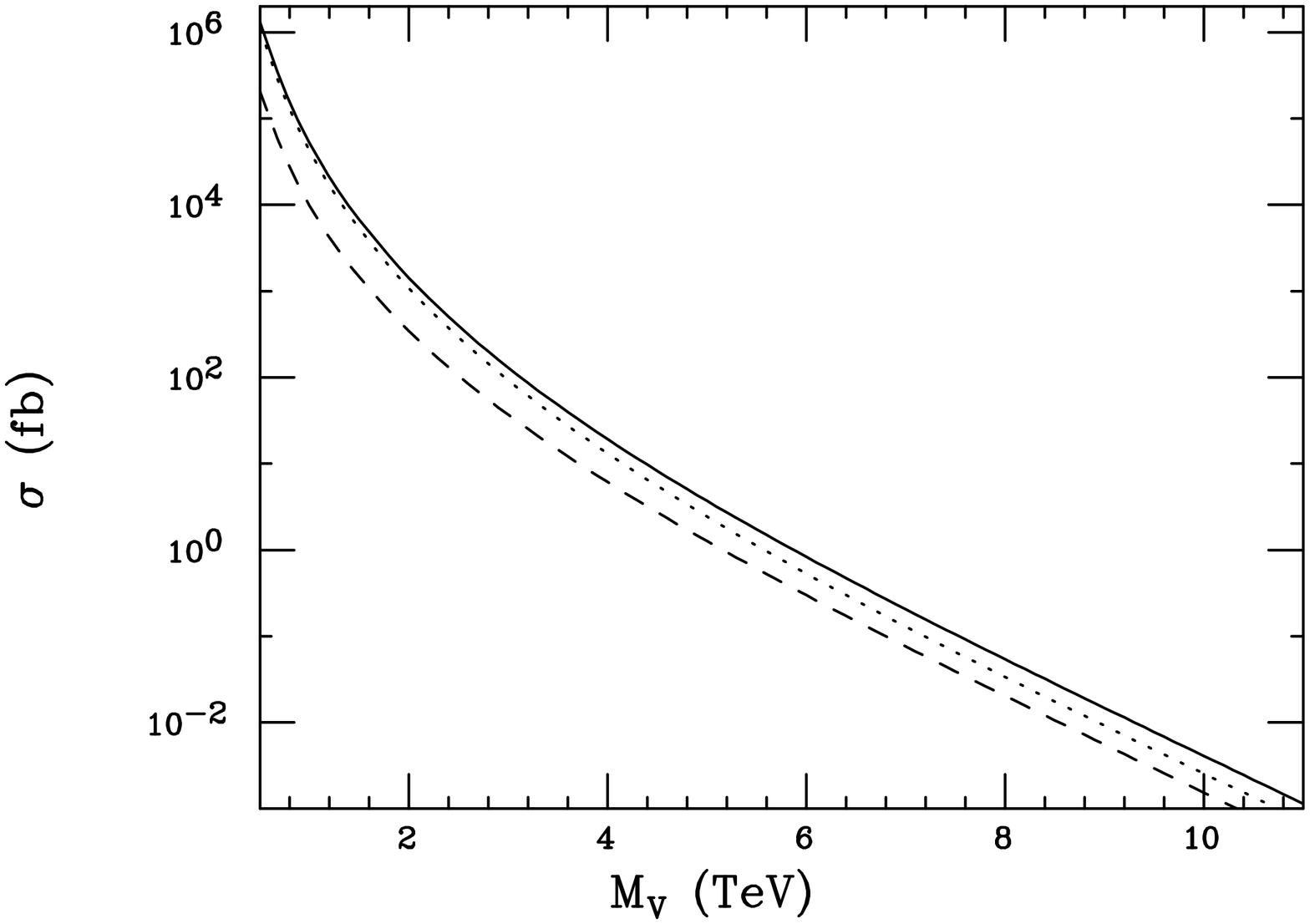,height=8.1cm,width=8.1cm,angle=-90}
\hspace*{-5mm}
\psfig{figure=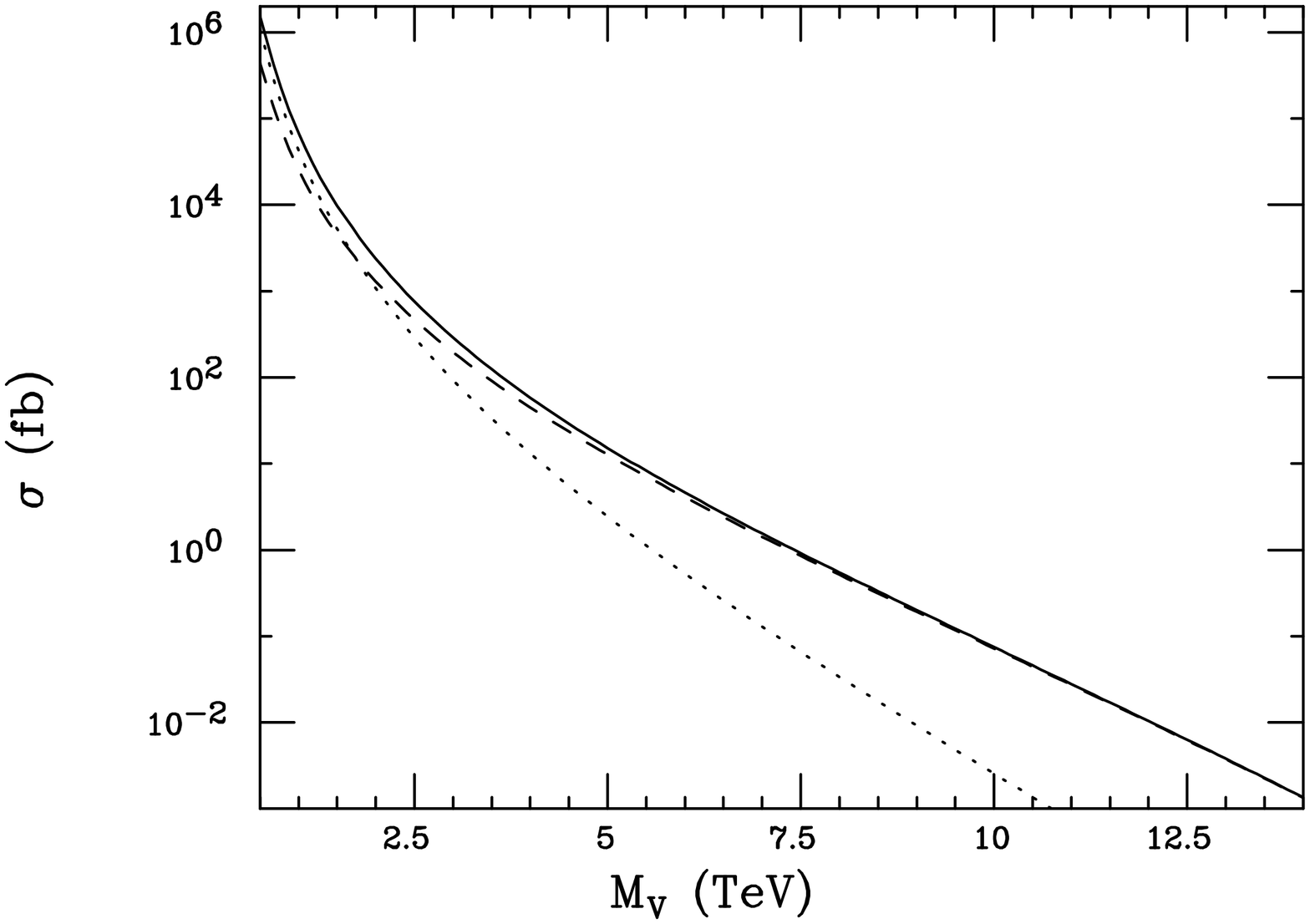,height=8.1cm,width=8.1cm,angle=-90}}
\vspace*{-0.6cm}
\caption{Same as the previous figure but now for a spin-1 vector leptoquark 
with $\kappa=1$ (from Ref.~{\protect\cite{rizzoproc}}).}
\label{v60}
\end{figure}
\vspace*{0.1mm}

\begin{figure}[htbp]
\centerline{
\psfig{figure=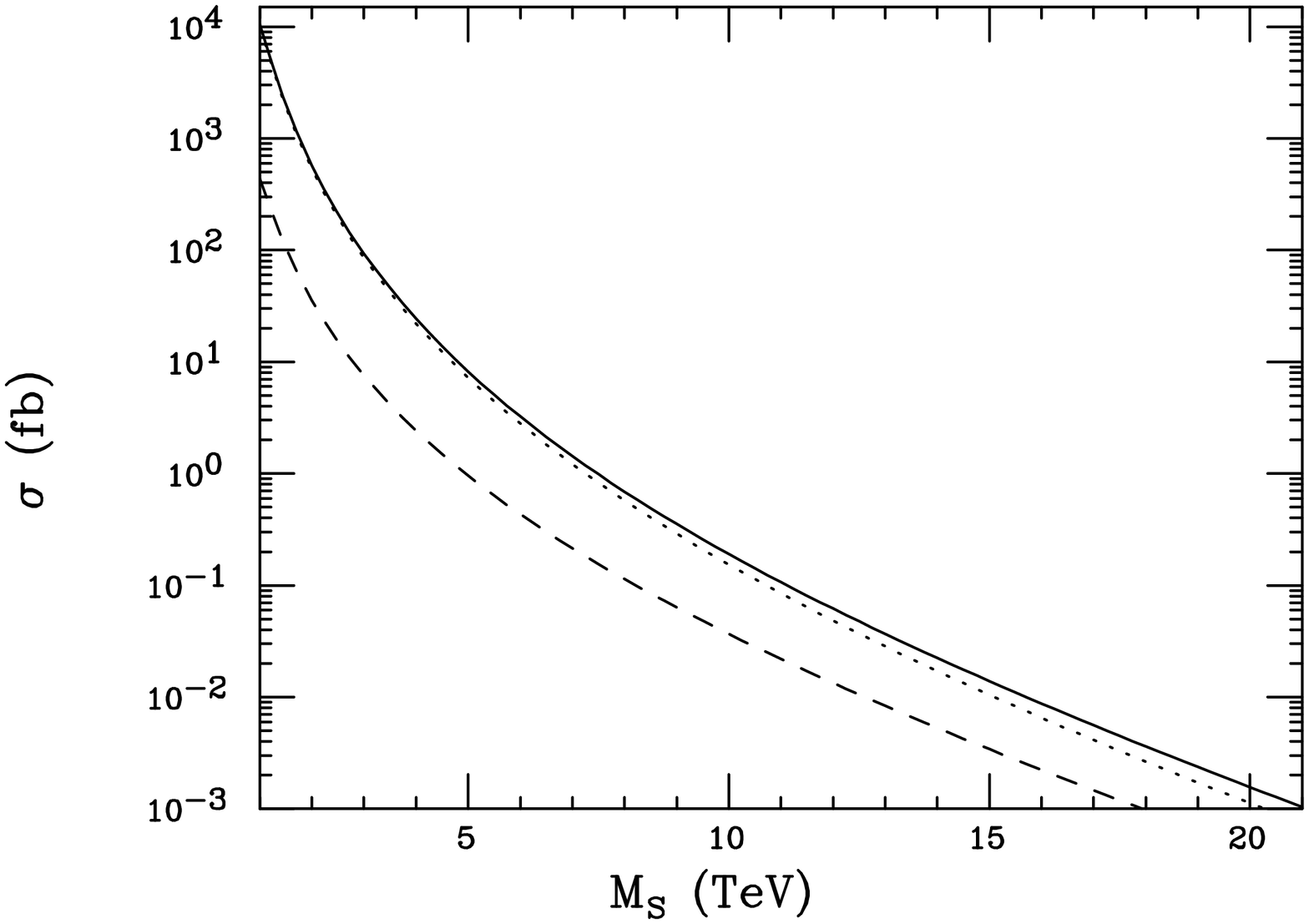,height=8.1cm,width=8.1cm,angle=-90}
\hspace*{-5mm}
\psfig{figure=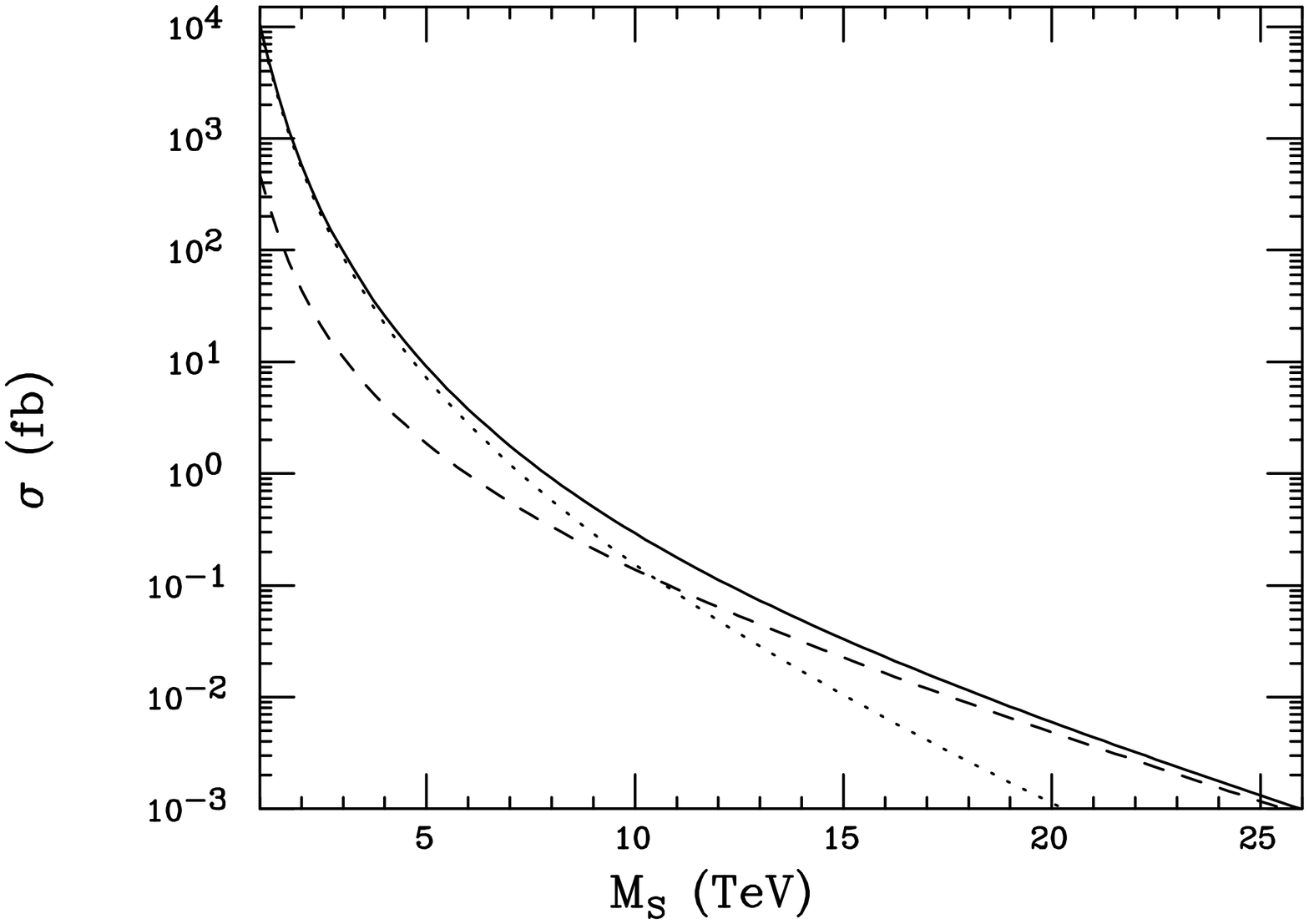,height=8.1cm,width=8.1cm,angle=-90}}
\vspace*{-0.6cm}
\caption{Same as Fig.1 but now at the 200 TeV PIPETRON collider (from 
Ref.~{\protect\cite{rizzoproc}}).}
\label{s200}
\end{figure}

\begin{figure}[htbp]
\centerline{
\psfig{figure=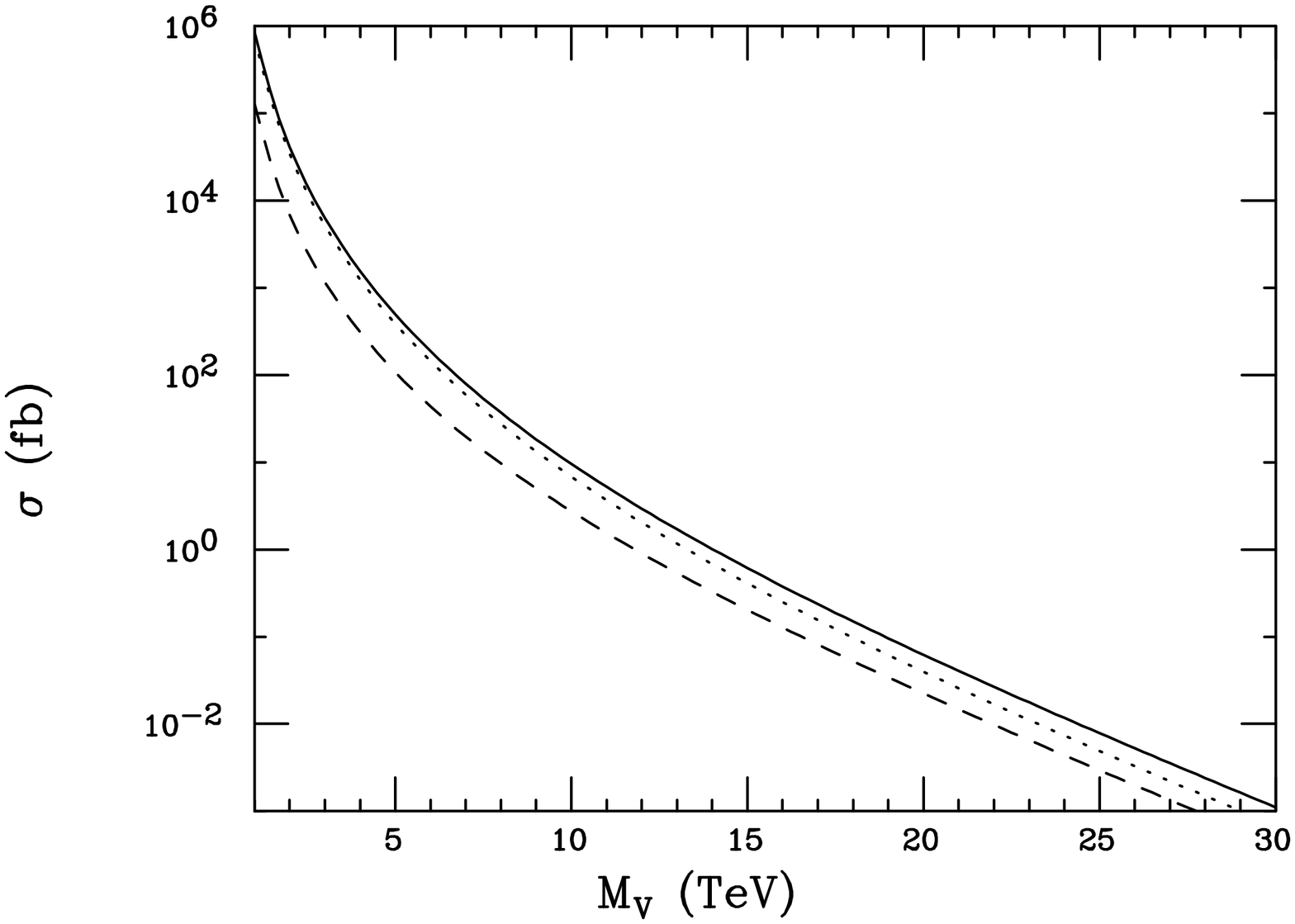,height=8.1cm,width=8.1cm,angle=-90}
\hspace*{-5mm}
\psfig{figure=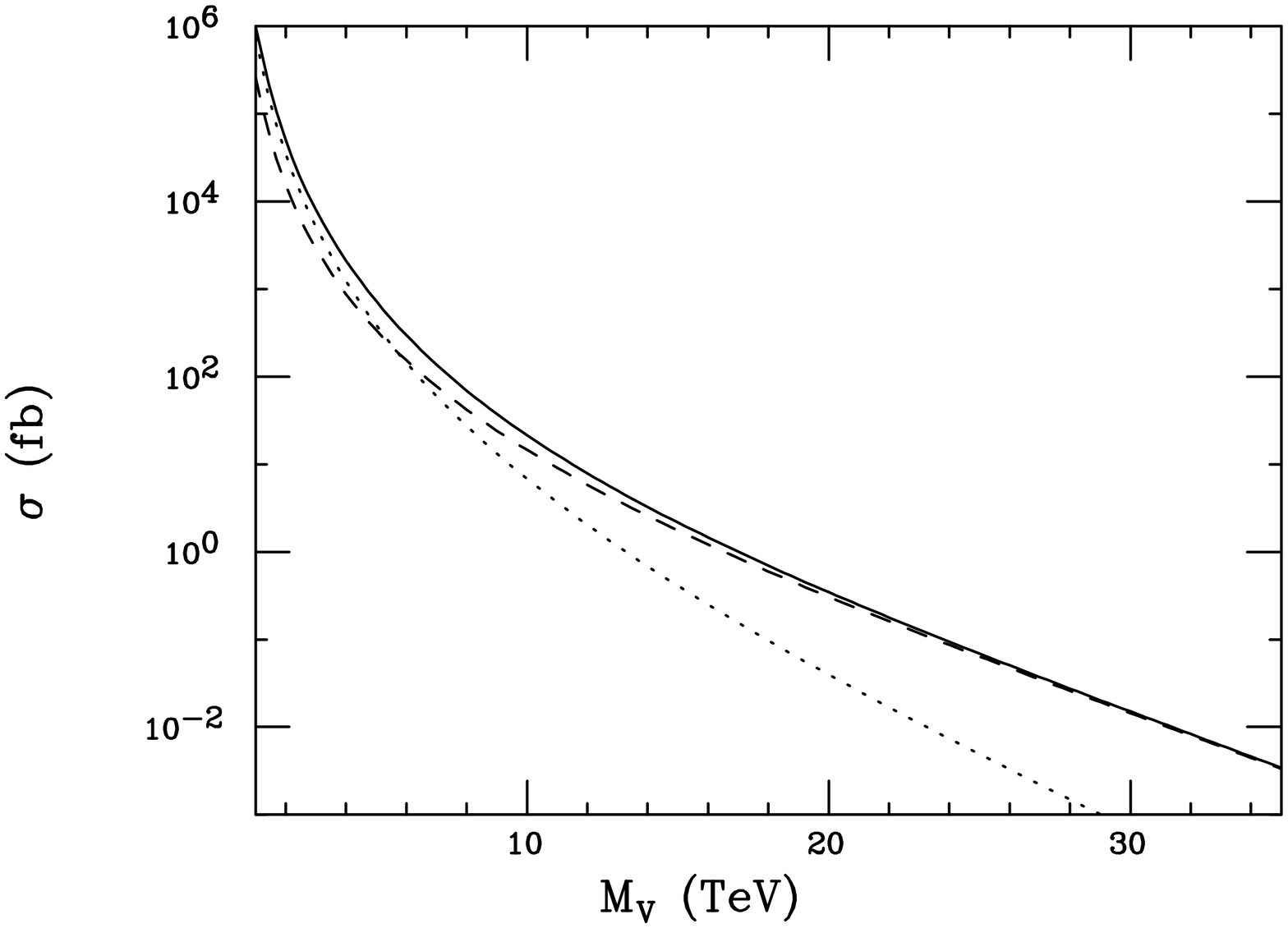,height=8.1cm,width=8.1cm,angle=-90}}
\vspace*{-0.6cm}
\caption{Same as Fig.2 but now for the 200 TeV PIPETRON collider (from 
Ref.~{\protect\cite{rizzoproc}}).}
\label{v200}
\end{figure}

The cross sections for $S$ and $V$ pair production at the $\sqrt s$=60 (LSGNA) 
and 200 (PIPETRON) TeV machines are displayed in 
Figures~\ref{s60},~\ref{v60},~\ref{s200} and~\ref{v200}.
The corresponding results for the Tevatron and LHC have been presented 
previously in, e.g., 
Ref.{\cite {review}}. The contributions of the 
subprocesses $gg\to SS,VV$ and $q\bar q \to SS,VV$ are displayed along with 
the total cross section.
The following conclusions can be drawn\cite{rizzoproc}:
\begin{itemize}
\item The vector leptoquark cross section is 
substantially larger than that for scalars in both $pp$ and $p\bar p$ 
collisions since the rates for both $gg\to VV$ and $q\bar q\to VV$ are larger 
than their scalar counterparts. 
\item Due to the 
contribution of the $q\bar q$ 
production mode, 
$p\bar p$ colliders have larger leptoquark cross sections than do $pp$ 
colliders.
\item  At $pp$ machines, for both vector and scalar leptoquarks, the cross 
sections are dominated by the $gg$ process out to the machine's anticipated 
mass reach.
\item In the $\sqrt s=60$ TeV $p\bar p $ case, the $q\bar q$ process 
dominates over $gg$ for masses 
greater than about 3.0(1.8) TeV for scalar(vector) leptoquarks.
In the $\sqrt s=200$ TeV $p\bar p $ case, the $q\bar q$ process 
dominates over $gg$ for masses 
greater than about 10(6) TeV for scalar(vector) leptoquarks.
\end{itemize}

\vspace*{-0.5cm}
\nn
\vspace*{0.1mm}
\vspace*{-0.5cm}
\nn
\vspace*{0.1mm}

Table~\ref{leptos} summarizes and compares the search reaches for both scalar 
and vector leptoquarks at the Tevatron and LHC as well as the hypothetical 
60 and 200 TeV $pp$ and $p\bar p$ colliders. Rizzo's results for the Tevatron 
confirm the expectations of the TeV2000 Study Group {\cite {tev2000}}, 
who also 
assume the 10 event discovery limit, while those obtained for 
the LHC are somewhat smaller{\cite {wrochnaproc}} than that given by the fast 
CMS detector simulation described in the next section. 

\begin{table}[htbp]
\centering
\begin{tabular}{lccc}
\hline
\hline
Machine  & ${\cal L}(fb^{-1})$  & $S$ & $V$ \\
\hline
LHC                    &  100  & 1.34(1.27)  & 2.1(2.0)    \\
60 TeV($pp$)           &  100  & 4.9(4.4)    & 7.6(7.0)    \\
60 TeV($p\bar p$)      &  100  & 5.7(5.2)    & 9.6(9.0)    \\
200 TeV($pp$)          &  1000 & 15.4(14.1)  & 24.2(23.3)  \\
200 TeV($p\bar p$)     &  1000 & 18.1(16.2)  & 31.1(29.0)  \\
TeV33                  &  30 & $\simeq 0.35$ & $\simeq 0.58$\\
\hline
\hline
\end{tabular}
\caption{Search reaches in TeV for scalar($S$) and vector($V$) leptoquarks at 
future hadron 
colliders assuming a branching fraction into a charged lepton plus a jet of 
unity($1/2$). For vector leptoquarks, $\kappa=1$ has been assumed and in both 
cases the MRSA$'$ parton densities have been employed. These results are 
based on the assumption of 10 signal events (from 
Ref.~{\protect\cite{rizzoproc}}).}
\label{leptos}
\end{table}

\section{Pair Production of Leptoquarks in the CMS Detector}

Wrochna\cite{wrochnaproc} carried out a study of the ability of the CMS 
detector to discover 
a second generation leptoquark using its
muon-jet decay. The CMS detector was simulated using the package 
CMSJET\cite{cmsjet}. 
The leptoquark, being a heavy particle, gives rise to 
harder muon and jet spectrums than the Standard Model backgrounds. 
Therefore a cut on the transverse 
momenta of the muons drastically reduces the 
background. Other cuts to isolate the signal are discussed in 
Ref.~\cite{wrochnaproc}.

The signal and background after imposition of all the kinematic and topological
cuts is shown in Fig.~5 for $100~{\rm fb}^{-1}$ of luminosity. The reach of 
the CMS detector in leptoquark mass is about 1.6~TeV, at which point  
the number of 
signal events becomes marginal.
\begin{center}
\vspace*{-3.0in}
\epsfxsize=4.2in
\hspace*{0in}
\epsffile{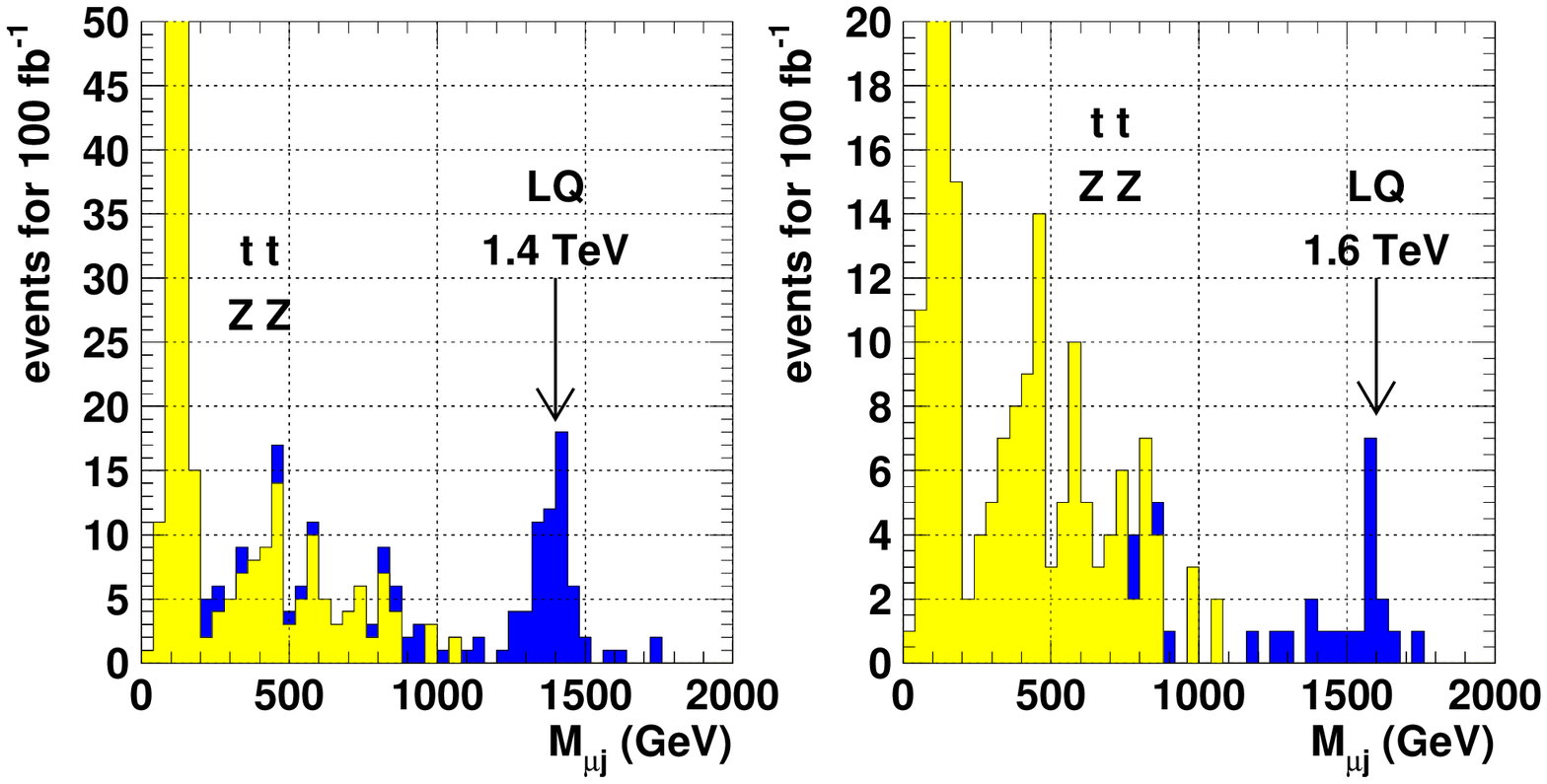}

\vspace*{0.5in}
\parbox{5.5in}{\small  Figure 5: Leptoquark signal and 
background mass distribution
in the CMS detector (from Ref.~{\protect\cite{wrochnaproc}}).}
\end{center}

\section{Single Leptoquark Production at Lepton Colliders}

Leptoquark production and identification was studied for lepton colliders
by Doncheski and Godfrey\cite{godfreyproc}. 
The production modes often considered are pair 
production for $e^+e^-$ and $\mu^+\mu^-$ machines. 
Single leptoquark production can be arise
for the $e\gamma $ mode (A muon beam cannot be converted into a 
photon beam for kinematic reasons\cite{sf}.). Leptoquarks can be produced 
singly in the $e\gamma $ mode, so higher masses can be probed than in 
double leptoquark production.
$e^+e^-$ scattering. Furthermore single leptoquark production can take place
at $e^+e^-$ and $\mu^+\mu^-$ machines by considering Weisacker-Williams photons
inside the incident leptons.
The production cross section depends on an unknown Yukawa coupling $g$; in 
the contribution of Doncheski and Godfrey 
this coupling is chosen to be equal to the 
electromagnetic coupling, i.e. $g^2/4\pi = \alpha _{em}$.
One can also use polarization and angular distributions 
to determine the properties of the leptoquarks. One useful observable that
has been defined\cite{ALLdef} to isolate the spin of a leptoquark is 
the double asymmetry
\begin{eqnarray}
A_{LL}={{(\sigma ^{++}+\sigma ^{--})-(\sigma ^{+-}+\sigma ^{-+})}
\over {(\sigma ^{++}+\sigma ^{--})+(\sigma ^{+-}+\sigma ^{-+})}}\;,
\end{eqnarray}
where the first index is the final state electron helicity and the second 
index is the final state quark helicity. Scalar leptoquarks 
only contribute when the 
electron and quark helicities are the same, and vector leptoquarks only 
contribute when they are opposite. Therefore at the {\it parton} level one has 
the asymmetry ${\hat a}_{LL}^{}=\pm 1$ for scalars and vectors, and one expects
this division to survive the folding in of the parton distribution functions.

A second observable for identifying leptoquarks is the left-right asymmetry, 
defined as 
\begin{eqnarray}
A^{+-}={{\sigma ^{+}-\sigma ^{-}}
\over {\sigma ^{+}+\sigma ^{-}}}={{C_L^2-C_R^2}\over {C_L^2+C_R^2}}\;.
\end{eqnarray}
This measurement can be used to determine the chirality of the leptoquark
coupling.

The high energy photon is obtained in one of two ways:
(1) as a Weisacker-Williams photon, or more optimistically (2) as a 
backscattered laser photon. The resulting photon is then resolved into its
hadronic content as shown in Fig.~6 and single leptoquark production can 
result. The backscattered photon gives a slightly reduced 
maximum center-of-mass energy than does a Weisacker-Williams photon, 
but it gives 
a harder photon spectrum with higher luminosity, but requires including the
backscattering option in the collider design. The cross sections for single 
leptoquark production are shown in Figs.~7 and 8. 
The cross section is significantly 
higher for the backscattered photon, but the ultimate reach in energy is 
slightly less.

\begin{center}
\vspace*{-1.3in}
\epsfxsize=3.0in
\hspace*{0in}
\epsffile{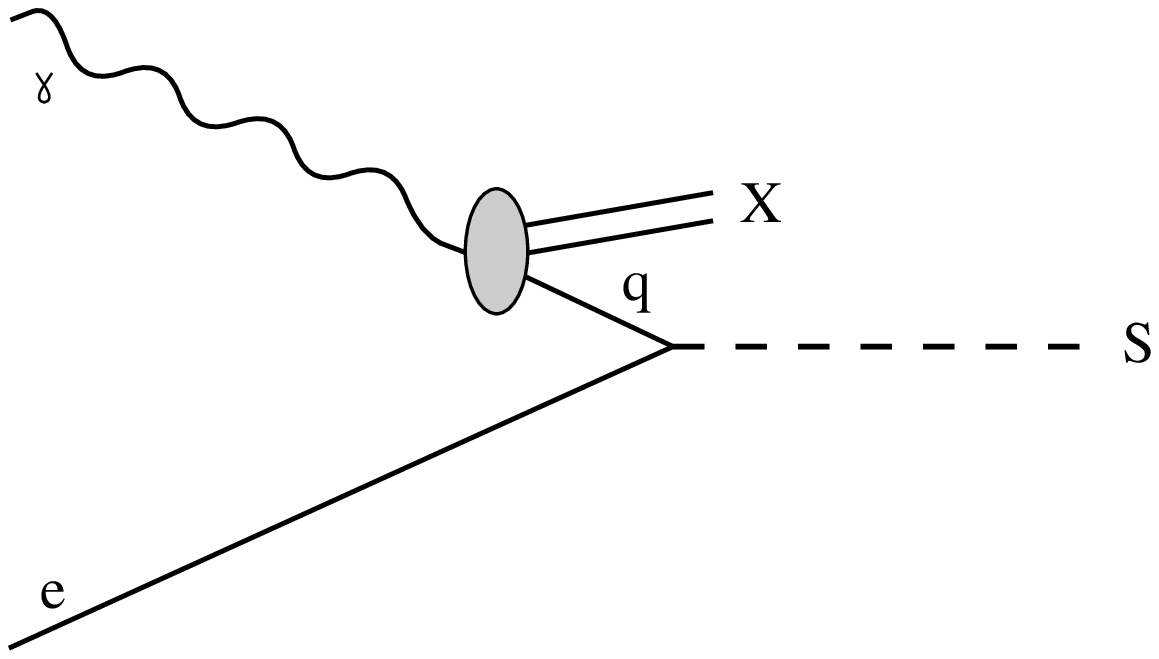}

\vspace*{-1.1in}
\parbox{5.5in}{\small  Figure 6: The resolved photon contribution for leptoquark 
production in $e\gamma $ collisions (from Ref.~\cite{godfreyproc}).}
\end{center}

\begin{center}
\vspace*{-1.6in}
\epsfxsize=3.6in
\hspace*{-0.3in}
\epsffile{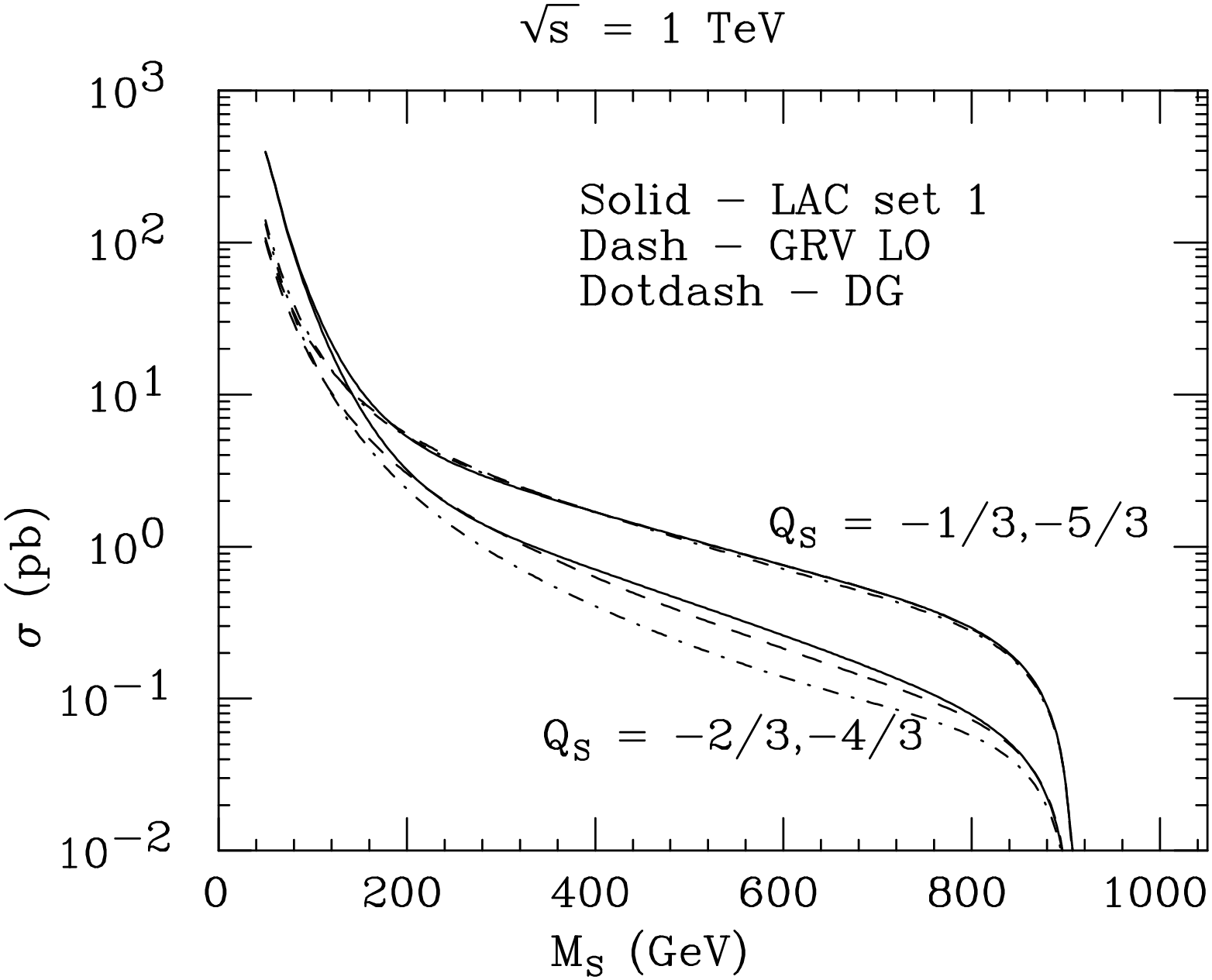}

\vspace*{0.3in}
\parbox{5.5in}{\small  Figure 7: The cross sections for leptoquark 
production due 
to resolved photon contributions in $e\gamma$ collisions for laser 
backscattered photons at a $\sqrt{s}=1~TeV$ collider. The solid, dashed,
dot-dashed lines are for resolved photon distribution functions LAC, GRV and
DG respectively (from Ref.~\cite{godfreyproc}).}

\end{center}

\begin{center}
\vspace*{-0.8in}
\epsfxsize=3.6in
\hspace*{-0.3in}
\epsffile{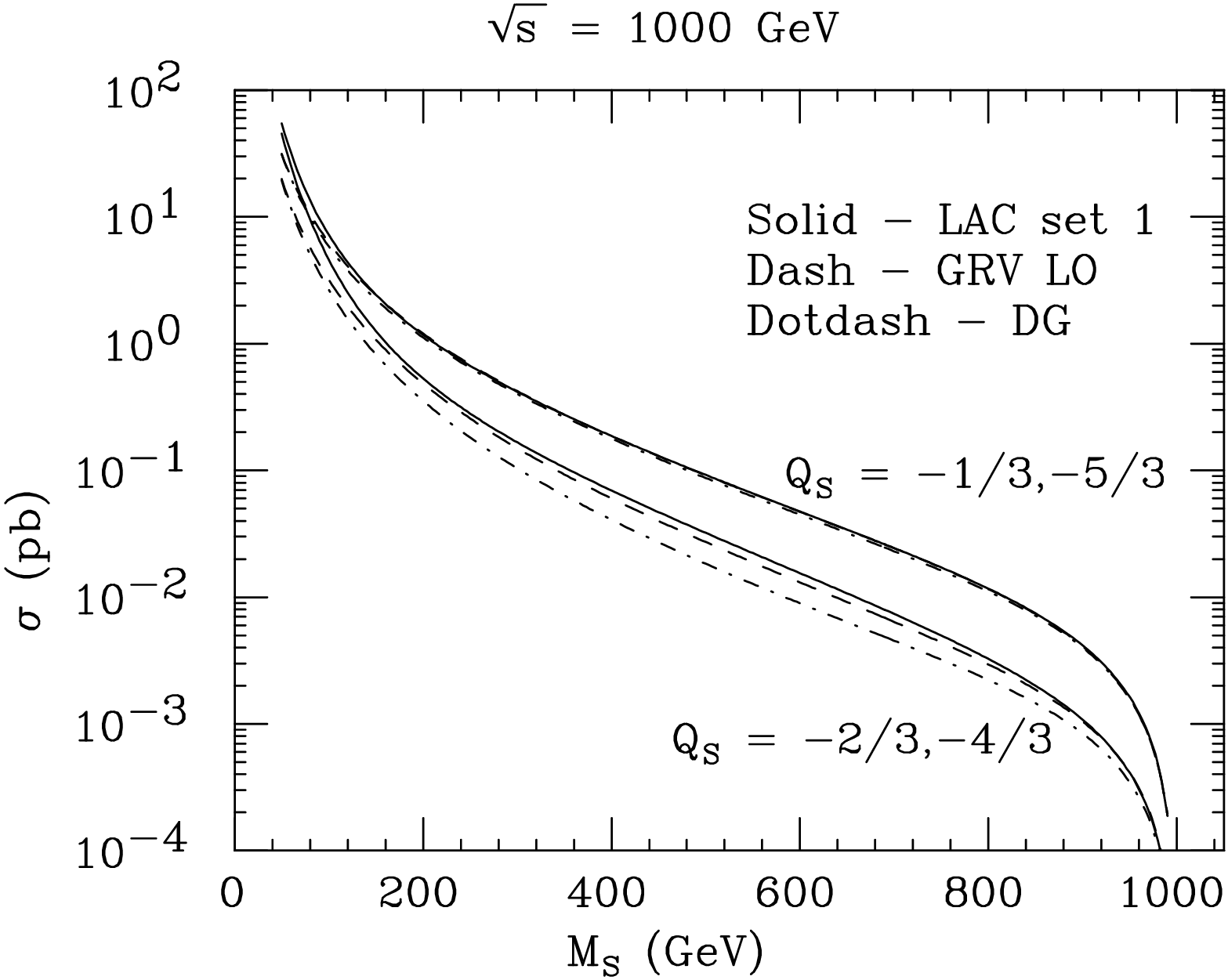}

\vspace*{0.3in}
\parbox{5.5in}{\small  Figure 8: The cross sections for leptoquark 
production due 
to resolved photon contributions in $e\gamma$ collisions for 
Weizs\"acker-Williams photons at a $\sqrt{s}=1~TeV$ collider. 
The solid, dashed,
dot-dashed lines are for resolved photon distribution functions LAC, GRV and
DG respectively (from Ref.~\cite{godfreyproc}).}
\end{center}

For colliders with a center-of-mass energies above 1~TeV, the reach for single
leptoquark production in this process is essentially the kinematic limit 
provided the planned luminosities of the machines is indeed realized. For
a $\sqrt{s}=500$~GeV machine there are some small differences between 
$e^+e^-$ machines and $\mu^+\mu^-$ machines: there are larger $u$ and $d$ 
content in the photon because their mass is smaller, so an $e^+e^-$ collider
have a 25\% higher reach than a $\mu^+\mu^-$ collider at the same energy.
However, it should be remembered that the two machines are probing different
leptoquarks (first-generation versus second-generation), so that the searches
are actually complementary.
The discovery limits obtained by Donchecki and Godfrey\cite{godfreyproc} are 
given in Table~\ref{godfreytab}.

\begin{table}[godfreytab]
\centering
\begin{tabular}{lrrrrr}
\hline
\hline
& & $e^+e^-$ Colliders & & & \\
\hline
$\sqrt{s}$~(TeV)   & $L(fb^{-1})$  & Scalar &  & Vector & \\
\hline
                 & & -1/3, -5/3 & -4/3, -2/3 & -1/3, -5/3 & -4/3, -2/3 \\
\hline 
0.5 & 50   &  490  &  470 &  490  &  480 \\
1.0 & 200  &  980  &  940 &  980  &  970 \\
1.5 & 200  & 1440  & 1340 & 1470  & 1410 \\
5.0 & 1000 & 4700  & 4200 & 4800  & 4500 \\
\hline
\hline
& & & & & \\
& & $e\gamma $ Colliders & & & \\
\hline
$\sqrt{s}$~(TeV)   & $L(fb^{-1})$  & Scalar &  & Vector & \\
\hline
                 & & -1/3, -5/3 & -4/3, -2/3 & -1/3, -5/3 & -4/3, -2/3 \\
\hline 
0.5 & 50   &  450  &  450 &  450  &  440 \\
1.0 & 200  &  900  &  900 &  910  &  910 \\
1.5 & 200  & 1360  & 1360 & 1360  & 1360 \\
5.0 & 1000 & 4500  & 4400 & 4500  & 4500 \\
\hline
\hline
& & & & & \\
& & $\mu^+\mu^-$ Colliders & & & \\
\hline
$\sqrt{s}$~(TeV)   & $L(fb^{-1})$  & Scalar &  & Vector & \\
\hline
                 & & -1/3, -5/3 & -4/3, -2/3 & -1/3, -5/3 & -4/3, -2/3 \\
0.5 & 0.7   &  250  &  170 &  310  &  220 \\
0.5 & 50  &  400  &  310 &  440  &  360 \\
5.0 & 1000 & 3600  & 3000 & 3700  & 3400 \\
\hline
\hline
\end{tabular}
\caption{Search reaches in TeV for scalar($S$) and vector($V$) leptoquarks at 
future hadron 
colliders assuming a branching fraction into a charged lepton plus a jet of 
unity($1/2$). For vector leptoquarks, couplings of electromagnetic strength
have been assumed and in both 
cases the MRSA$'$ parton densities have been employed. These results are 
based on the assumption of 10 signal events 
(from Ref.~{\protect\cite{godfreyproc}}).}
\label{godfreytab}
\end{table}

\section{Indirect Searches for Leptoquarks}

At $e^+e^-$ and $\mu^+\mu^-$ colliders, pairs of leptoquarks can be produced 
directly via the $s$-channel $\gamma $ and $Z$ exchange. The reach for the 
leptoquark mass for this mode is essentially the kinematic limit, i.e.
$M_S< \sqrt{s}/2$. However even if a leptoquark is too massive to be produced 
directly, it can contribute\cite{hr,dreiner,choudhury} 
indirectly to the process 
$\ell^+\ell^-\to q\bar{q}$ by interfering with the Standard Model diagrams
as shown in Fig.~9. The leptoquark interacts via a Yukawa coupling which
can be parametrized in the form
\begin{eqnarray}
{\cal L}&=&gS\bar{q}(\lambda_L P_L + \lambda_R P_R)\ell \;,
\end{eqnarray}
where $g$ is the weak coupling constant (to set the overall magnitude of the 
interaction) and $\lambda _{L,R}$ are dimensionless constants. 
$P_L$ and $P_R$ are the
left- and right-handed projectors. The amplitudes for the diagrams presented 
in Fig.~9 have been presented for
the unpolarized case in Ref.~\cite{hr}, and is generalized to the case with
polarization in Ref.~\cite{choudhury}. The size of the interference effect 
is determined by the three parameters $M_S$, $\lambda_L$ and $\lambda_R$.

By examining the overall rate and the angular distribution,
indirect evidence for leptoquarks can be obtained. Berger\cite{bergerproc} 
examined
the bounds which can be placed on the leptoquark mass including the option 
of polarizing the electron and muon beams.
The polarization of the beams of a lepton collider can serve two purposes
in indirect leptoquark searches: 
(1) it can extend the reach of the indirect search by serving to enhance 
the fraction of initial leptons to which the 
leptoquark couples; (2) it can measure the left-handed and right-handed 
couplings of the leptoquark separately.

\begin{center}
\epsfxsize=4.5in
\hspace*{0in}
\epsffile{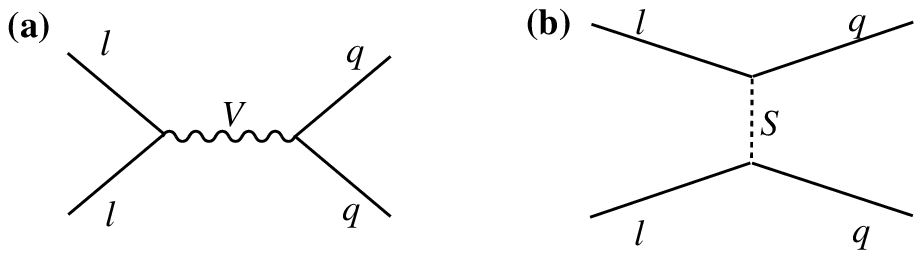}

\vspace*{0in}
\parbox{5.5in}{\small  Figure 9: The Feynman diagrams for the process 
$\ell^+\ell^-\to q\overline{q}$ include the (a) Standard Model diagrams
involving $s$-channel $V=\gamma,Z$ exchange,
and (b) the hypothetical $t$-channel leptoquark $S$ exchange 
(from Ref.~{\protect\cite{bergerproc}}).}
\end{center}

The deviations from the Standard Model appear in the total cross section and 
the angular distribution of the produced quarks\cite{hr}. 
The total cross section and the forward-backward asymmetry, $A_{FB}$, amount
to integrating this distribution in one or two bins respectively. The 
statistical significance of the signal is determined by calculating a $\chi^2$ 
for the deviation from the expectation in the Standard Model\cite{choudhury},
\begin{eqnarray}
\chi ^2&=&\sum _{j=1}^{18}{{(n_j^{\rm LQ}-n_j^{\rm SM})^2}
\over {n_j^{\rm SM}}}\;,
\end{eqnarray}
where $n_j^{\rm SM}$ is the number of 
events expected in each $\Delta \cos \theta =0.1$ bin in the Standard Model,
and $n_j^{\rm LQ}$ is the number of events including the leptoquark.

The additional piece in the Lagrangian that is of relevance to us can be
parametrized in the form
\begin{eqnarray}
{\cal L}&=&gS\bar{q}(\lambda_L P_L + \lambda_R P_R)\ell \;,
\end{eqnarray}
where $g$ is the weak coupling constant (to set the overall magnitude of the 
interaction) and $\lambda _{L,R}$ are dimensionless constants. $P_L$ and $P_R$ are the
left- and right-handed projectors. The size of the interference effect will be
determined by the three parameters $M_S$, $\lambda_L$ and $\lambda_R$.

Figure~10 
shows the 95\% c.l. bounds that could be achieved on a leptoquark with
right-handed couplings ($\lambda _L=0$) at a
$\sqrt{s}=4$~TeV $e^+e^-$ collider, with nonpolarized beams and with 
80\% and 100\% polarization of the electron beam. We have assumed 
integrated luminosity $L_0$  and efficiency $\epsilon$ for detecting the final 
state quarks so that $\epsilon L_0=70 {\rm fb}^{-1}$.
Polarization from 80\% to 100\% roughly brackets
the range that might reasonably be achievable for the electron beam. 
Figure~11 shows the same bounds for the case where the leptoquark has 
left-handed couplings ($\lambda _R=0$). 

In a muon collider both $\mu^+$ and $\mu^-$ beams can be at least partially
polarized, but perhaps with some loss of
luminosity\cite{feas}. If one
tolerates a drop in luminosity of a factor two, then one can achieve
polarization of both beams at the level of $P^-=P^+=34\%$.
It might be possible to maintain the luminosity at its full unpolarized value
if the proton source intensity (a proton beam is used to create pions that
decay into muons for the collider) could be increased\cite{feas}.
Results for each of these three possible scenarios
below in Fig.~12 for a leptoquark with right-handed couplings
and in Fig.~13 for
a leptoquark with left-handed couplings. In the former case polarization is
useful for improving the leptoquark bounds even with a loss of two in
luminosity.

\begin{center}
\epsfxsize=3.0in
\hspace*{0in}
\epsffile{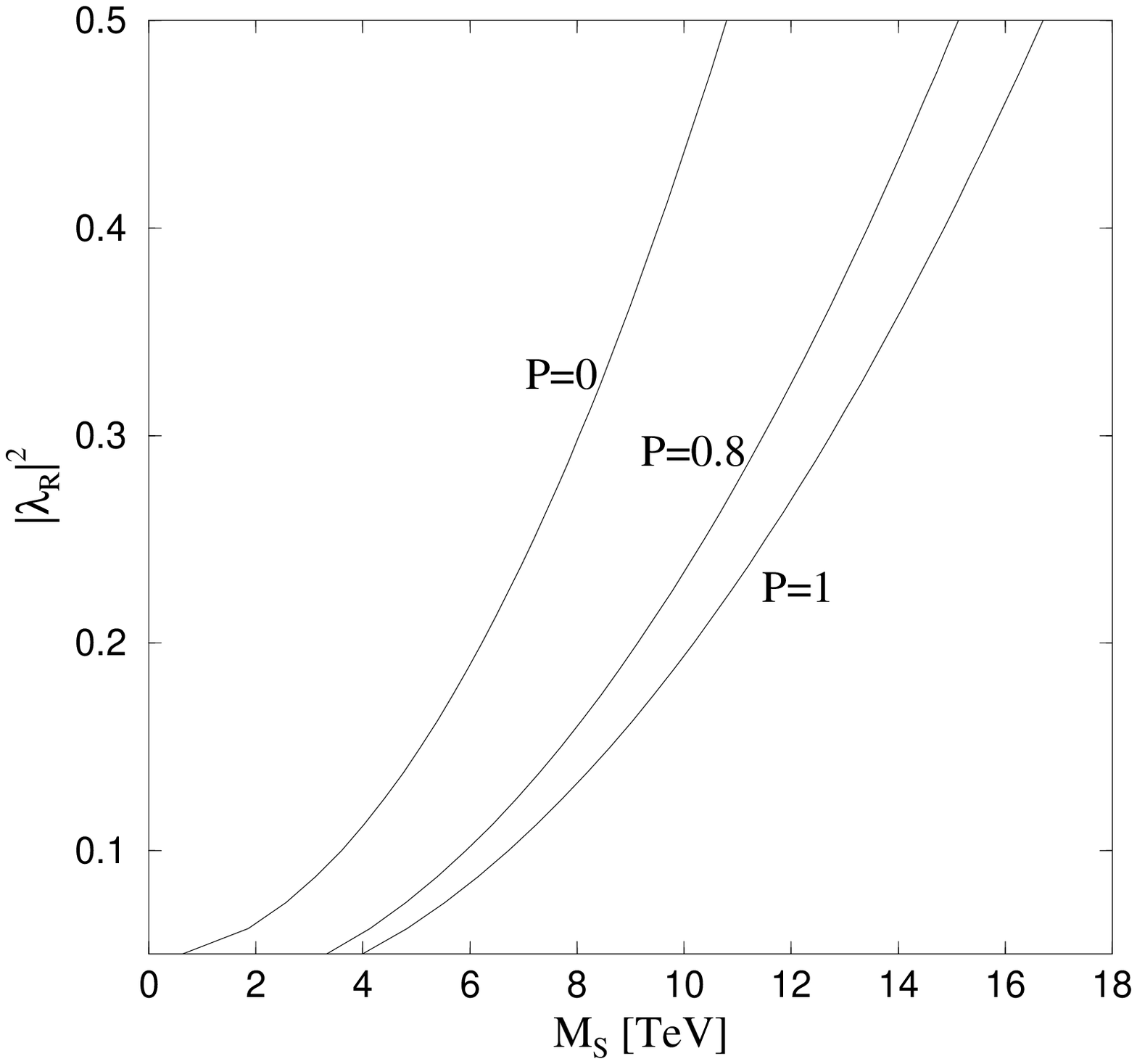}

\vspace*{-1.3in}
\parbox{5.5in}{\small  Figure 10: The 95\% c.l. bounds on leptoquark mass and 
couplings at 
a $\sqrt{s}=4$~TeV $e^+e^-$ collider for a leptoquark with right-handed 
couplings only ($\lambda _L=0$). The electron polarization $P$ is set to
0\%, 80\% and 100\%, and the positron is always unpolarized.
The area above each curve would be excluded 
(from Ref.~{\protect\cite{bergerproc}}).}
\end{center}

\begin{center}
\epsfxsize=3.0in
\hspace*{0in}
\epsffile{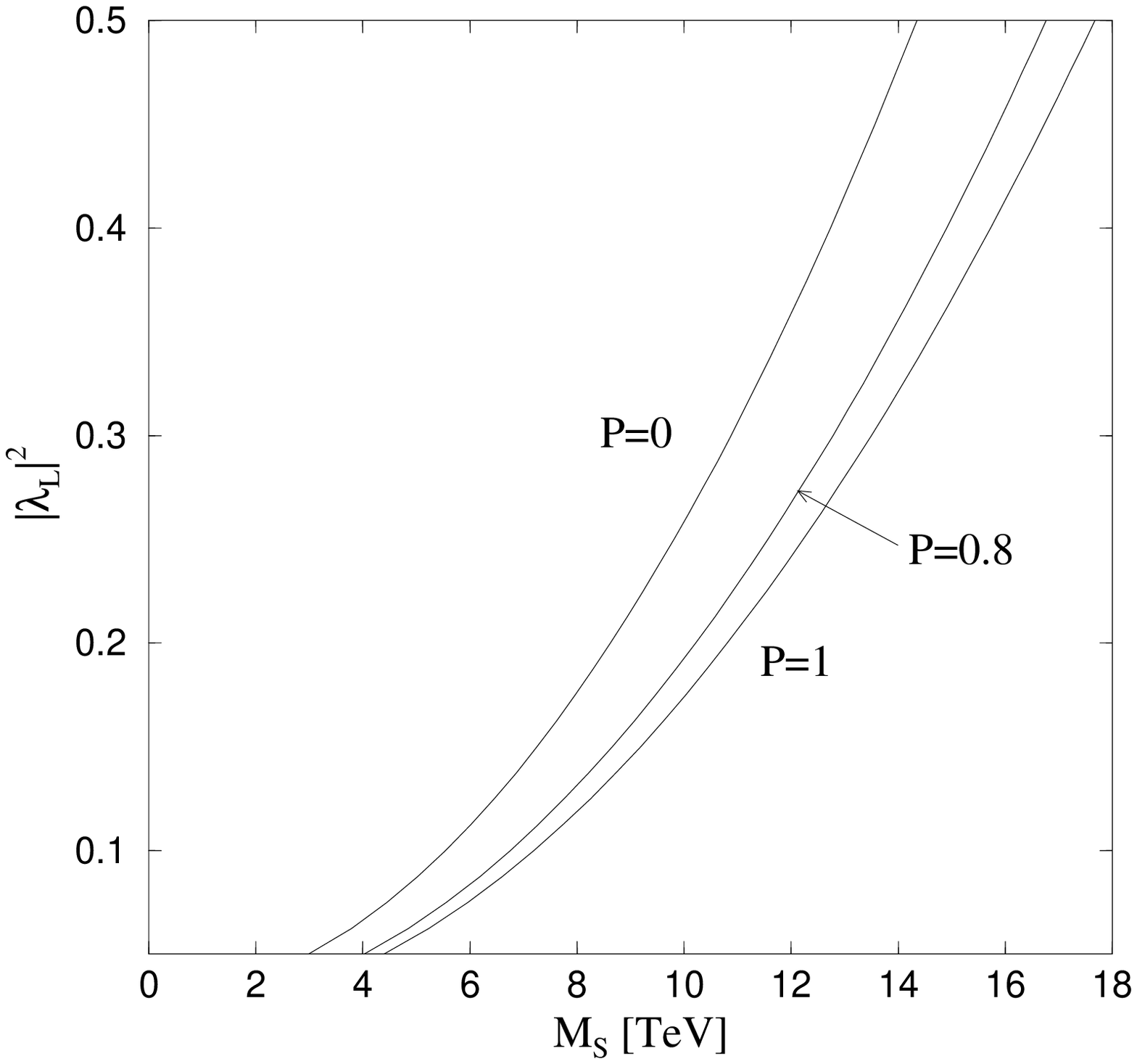}

\vspace*{-1.3in}
\parbox{5.5in}{\small  Figure 11: The 95\% c.l. bounds on leptoquark mass and 
couplings at 
a $\sqrt{s}=4$~TeV $e^+e^-$ collider for a leptoquark with left-handed 
couplings only ($\lambda _R=0$). The electron polarization $P$ is set to
0\%, 80\% and 100\%, and the positron is always unpolarized.
The area above each curve would be excluded 
(from Ref.~{\protect\cite{bergerproc}}).}
\end{center}

\begin{center}
\epsfxsize=3.0in
\hspace*{0in}
\epsffile{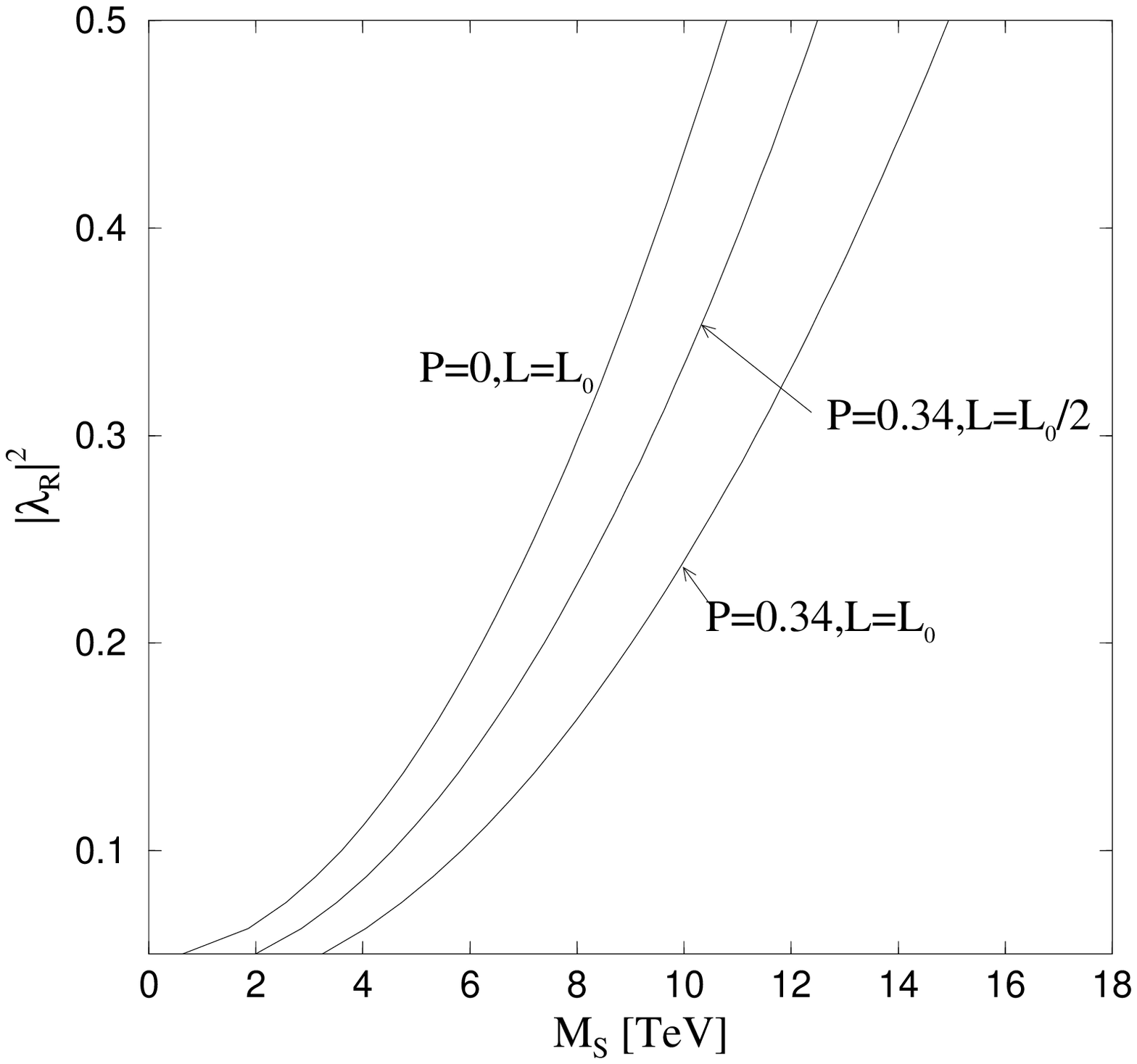}

\vspace*{-1.3in}
\parbox{5.5in}{\small  Figure 12: The 95\% c.l. bounds on leptoquark mass and 
couplings at 
a $\sqrt{s}=4$~TeV $\mu^+\mu^-$ collider for a leptoquark with right-handed 
couplings only ($\lambda _L=0$). The curves indicate the bounds for 
nonpolarized beams,  both $\mu^+$ and $\mu^-$ having 
polarization $P$ is set to 34\% and no reduction in luminosity, and 
both $\mu^+$ and $\mu^-$ having 
polarization $P$ is set to 34\% and a reduction in luminosity of 
a factor of two.
The area above each curve would be excluded 
(from Ref.~{\protect\cite{bergerproc}}).}
\end{center}

\begin{center}
\epsfxsize=3.0in
\hspace*{0in}
\epsffile{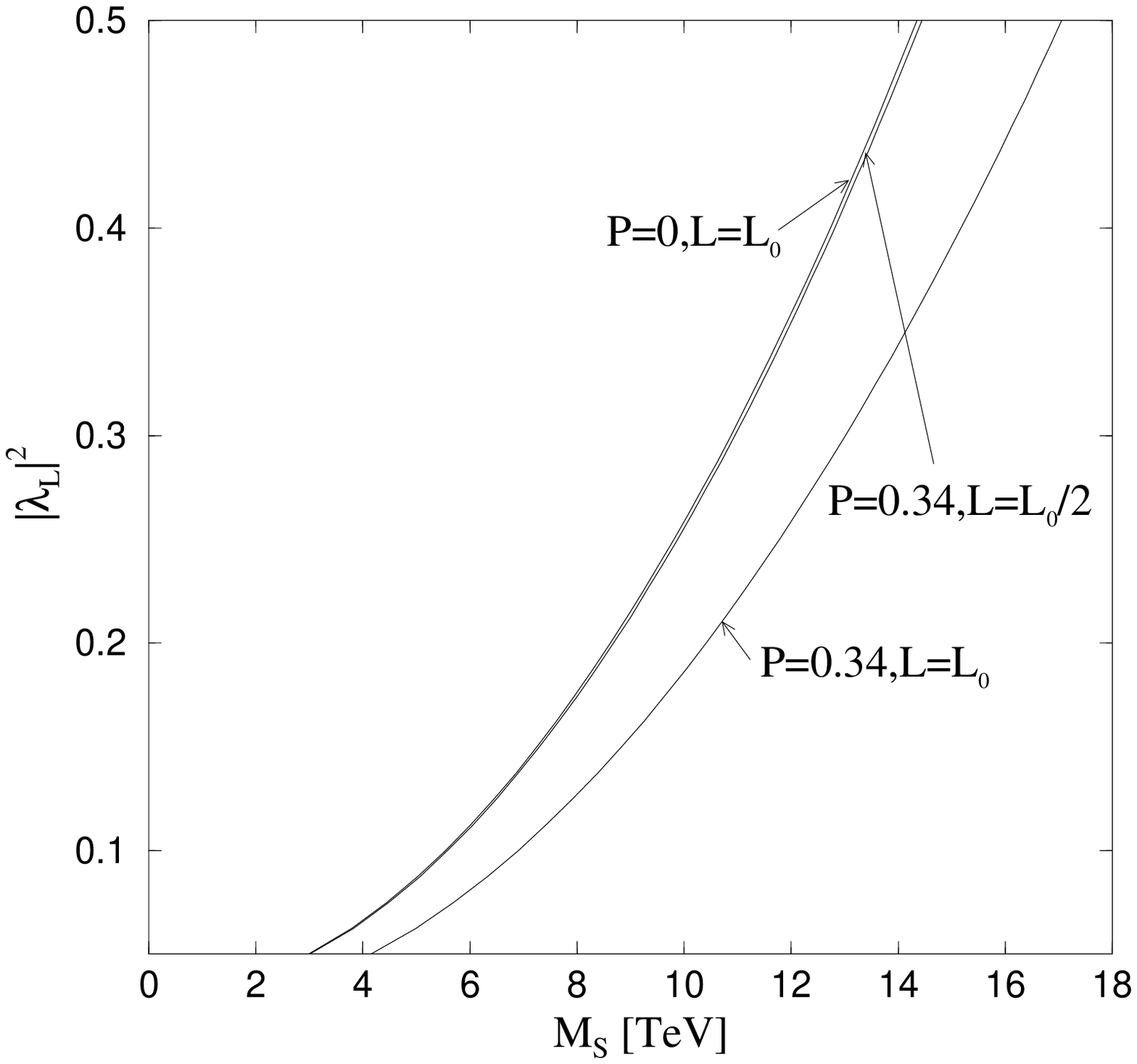}

\vspace*{-1.3in}
\parbox{5.5in}{\small  Figure 13: The 95\% c.l. bounds on leptoquark mass and 
couplings at 
a $\sqrt{s}=4$~TeV $\mu^+\mu^-$ collider for a leptoquark with left-handed 
couplings only ($\lambda _R=0$). The curves indicate the bounds for 
nonpolarized beams,  both $\mu^+$ and $\mu^-$ having 
polarization $P$ is set to 34\% and no reduction in luminosity, and 
both $\mu^+$ and $\mu^-$ having 
polarization $P$ is set to 34\% and a reduction in luminosity of 
a factor of two.
The area above each curve would be excluded
(from Ref.~{\protect\cite{bergerproc}}).}
\end{center}

One can compare the utility of polarizing both beams as opposed to 
polarizing just one beam. This can be done by comparing Figs.~10 and 12 
for the 
right-handed leptoquark case and Figs.~11 and 13 for the 
left-handed leptoquark case. The bounds for leptoquarks with 
interactions of order the weak coupling strength are summarized in Table III, 
for both 
left-handed couplings ($|\lambda _L|^2=0.5,|\lambda _R|^2=0$) and right-handed 
couplings ($|\lambda _R|^2=0.5,|\lambda _L|^2=0$). 
For both cases one sees that the 34\% polarization
of both beams gives roughly the same bounds as a collider with one beam 
polarized at the 80-90\% level. 

%%%%% A Table
%%
\begin{table}[h]
\begin{center}
\caption{Bounds on leptoquark masses at 98.6\% 
confidence level, assuming either
left-handed couplings ($|\lambda _L|^2=0.5,|\lambda _R|^2=0$) or
right-handed couplings ($|\lambda _L|^2=0,|\lambda _R|^2=0.5$)
(from Ref.~{\protect\cite{bergerproc}}).}
\label{tab:sample}
\begin{tabular}{l|c|c}
\hline
\hline
Luminosity and & & \\
Polarization($\ell^-,\ell^+$)  & Coupling & $M_S$-Bound (TeV)\\
\hline
$L_0$ (0\%,0\%) & Left & 14.3  \\
            &  Right & 10.8  \\
\hline
$L_0$ (80\%,0\%) &  Left & 16.8  \\
            &  Right & 15.1  \\
\hline
$L_0$ (100\%,0\%) &  Left & 17.7  \\
            &  Right & 16.7  \\
\hline
$L_0$ (34\%,34\%)  & Left & 17.1  \\
                          & Right & 14.9  \\
\hline
$L_0/2$ (34\%,34\%)  & Left & 14.4  \\
                          & Right & 12.5  \\
\hline
\hline
\end{tabular}
\end{center}
\end{table}
%%
%%%%%

\section{Bilepton Searches at the NLC}

A new particle which couples to two Standard Model leptons has been named a 
bilepton, and sometimes has been called a dilepton. Cuypers and 
Davidson\cite{cdproc} have studied the discovery prospects for the NLC.
They study all of the possible NLC modes, $e^+e^-$, $e^-e^-$, $e^-\gamma$ and
$\gamma \gamma$.
For their study they assumed a luminosity which scales with energy as
\begin{eqnarray}
{\cal L}{\rm [fb}^{-1}{\rm ]} &=& 200s{\rm [TeV}^{2}{\rm ]}\;, 
\end{eqnarray}
in the $e^+e^-$ mode. 
For the $e^-e^-$ mode, they assume the luminosity is reduced by a factor of 
two compared to the $e^+e^-$ mode.

The quantum numbers and couplings of the bileptons is shown in Table~\ref{tqn}.
The numerical index indicates bileptons which are singlets, doublets or 
triplets under the weak SU(2) gauge symmetry, and vector bileptons also 
carry an index $\mu$. Bileptons carry either zero total lepton number (and
some of the Standard Model particles fall into this category), or carry
two units of lepton number.
See Ref.~\cite{cdproc} for more details about their interactions.

If the NLC is operated in the $e^-e^-$ mode, it can produce doubly charged 
bileptons in the $s$-channel. The signal is a pair of like-sign leptons;
flavor violating processes like $e^-e^-\to \mu^-\mu^- $ might even be possible.
IF one knows the mass of the new particle, one could try to set the 
center of mass energy so as to sit on the resonance.
If the leptonic couplings are small,
it is possible that the signal will be reduced because the bilepton width 
is even smaller
than the beam energy spread.
Cuypers and Davidson find the the 95\% c.l. lower limits on the bilepton masses
are of the order of
\begin{eqnarray}
m_L^{}\gsim \sqrt{s}\times 50\lambda_{ee}\;,
\end{eqnarray}
where $\lambda_{ee}$ is the coupling of the bilepton to $e^-e^-$.

Singly-charged bileptons can be produced in $e^-\gamma$ scattering, but there
is no resonance in this case. Figure~14 shows the discovery 
potential found by Cuypers and Davidson  of 
$e^-\gamma $ collisions for several center of mass energies.

Singly-charged bileptons can also be pair produced in $e^+e^-$ collisions.
In Fig.~15 the number of expected
bilepton events is shown for right-polarized $e^+e^-$ collisions (the beams
are polarized to eliminate Standard Model backgrounds). A viable signal is
obtained all the way to the kinematic limit.

Above the kinematic limit ($m_L>\sqrt{s}/2$), 
the bileptons can have an indirect
effect on Bhahba scattering and produce significant deviations from the 
Standard Model predictions (this is analoqous to the indirect effects of 
leptoquarks described in the previous section). 
Cuypers and Davidson again find 
the 95\% C.L. lower limits on the 
bilepton masses
are of the order of
\begin{eqnarray}
m_L^{}\gsim \sqrt{s}\times 50\lambda_{ee}\;.
\end{eqnarray}

\vspace*{0in}
\begin{figure}[h]
\input{eg2nl.pstex}

\vspace{0.2in}
\small{Figure 14:
Smallest observable scalar \dil\ $L_1^{-}$
couplings to leptons
at the one standard deviation level
as a function of the \dil\ mass and coupling
in \ep\ collisions.
The collider's \pe\ \cm\ energies
are .5, 1, 2, and 3 TeV
from left to right (from Ref.~{\protect\cite{cdproc}}).
}
\label{feg2nl}
\end{figure}

\vspace*{0in}
\begin{figure}[h]
\input{pe2ll.pstex}

\vspace{0.2in}
\small{Figure 15:
Mass dependence of the number of pair-produced
singly-charged \dil s in \pe\ annihilations
(from Ref.~{\protect\cite{cdproc}}).
}
\label{fpe2ll}
\end{figure}

\newcommand{\sd}[1]{\raisebox{-2.5ex}[0ex][0ex]{$#1$}}
\renewcommand{\arraystretch}{2}

\begin{table}%[htb]
%
$
\setlength{\arraycolsep}{1em}
\begin{array}{||c||c|c|c|c|c|c|l|c|}
  \hline
  \hline
  & L & J & Y & T_3 & Q_\gamma & Q_Z  &
  \multicolumn{1}{c|}{\mbox{lepton couplings}}
  & \mbox{familiar sibling} \\
  \hline
  \hline
L_1^\mu & 0 & 1 & 0 & 0 & 0 & 0 & \bar e_Le_L ~(g_1) 
  \quad \bar\nu_L\nu_L ~(g_1) & \gamma \quad Z^0 \quad Z' \\
  \hline
\tilde L_1^\mu & 0 & 1 & 0 & 0 & 0 & 0 & 
  \bar e_Re_R ~(\tilde g_1) & \gamma \quad Z^0 \quad Z' \\
  \hline
\sd{L_2} & \sd{0} & \sd{0} & \sd{1/2} & 1/2 & 1 &
 - {2\swt-1\over2\sw\cw} & \bar\nu_Le_R ~(g_2) & H^+ \\
  &&&& -1/2 & 0 &- {1\over2\sw\cw} & \bar e_Le_R ~(g_2) & H \\
  \hline
&&&& 1 & 1 & {\cw\over\sw} & \bar\nu_Le_L ~(\sqrt{2}g_3) 
  & W^+ \quad W'^+ \\
L_3^\mu & 0 & 1 & 0 & 0 & 0 & 0 & e_Le_L 
  ~(-g_3) \quad \bar\nu_L\nu_L ~(g_3) &\gamma \quad Z^0 \quad Z' \\
&&&& -1 &- 1 & -{\cw\over\sw} & \bar e_L\nu_L ~(\sqrt{2}g_3) 
  & W^- \quad W'^- \\
  \hline
L_1 & 2 & 0 & 1 & 0 & 1 & -{\sw\over\cw} & 
  e_L\nu_L ~(\lambda_1) ~(\mbox{antisymm.}) \\
  \cline{1-8}
\tilde L_1 & 2 & 0 & 2 & 0 & 2 & -2{\sw\over\cw} 
  & e_Re_R ~(\tilde\lambda_1) \quad(\mbox{symm.}) \\
  \cline{1-8}
  \sd{L_2^\mu} & \sd{2} & \sd{1} & \sd{3/2} & 1/2 & 2 &
   -{4\swt-1\over2\sw\cw} & e_Re_L ~(\lambda_2) \\
&&&& -1/2 & 1 & - {2\swt+1\over2\sw\cw} & e_R\nu_L ~(\lambda_2) \\
  \cline{1-8}
&&&& 1 & 2 & -{2\swt-1\over\sw\cw} & e_Le_L ~(\sqrt{2}\lambda_3) \\ 
L_3 & 2 & 0 & 1 & 0 & 1 & -{\sw\over\cw} 
  & e_L\nu_L~(\lambda_3) \qquad(\mbox{symm.}) \\
&&&& -1 & 0 & - {1\over\sw\cw} & \nu_L\nu_L ~(-\sqrt{2}\lambda_3) \\
  \cline{1-8}
\end{array}
$
%
\vspace{0cm}
\caption{
Major quantum numbers and couplings of the \dil s (from 
Ref.~{\protect\cite{cdproc}}).
}
\label{tqn}
\end{table}

\section{Neutral Heavy Leptons}

Kalyniak and Melo\cite{kalyniakproc} have studied  
the production of a single neutral heavy lepton (NHL) in 
association with a massless neutrino in $e^+e^-$ and $\mu^+\mu^-$ 
colliders\cite{buch,ad,mmv}. The models considered have two new weak 
isosignlet neutrino fields per generation yielding three massless neutrinos 
($\nu _i$) and three Dirac NHL's ($N_a$)\cite{bkm,mv,bsvmv,ww}.
The Feynman diagrams are shown in Fig.~16.
The weak interaction eigenstates, $\nu_\ell, \ell =e,\mu,\tau$, are related
to the neutrino mass eignestates via two $3\times 3$ mixing matrices:
\begin{eqnarray}
\nu_\ell &=&\sum _{i=1,2,3}(K_L)_{li}\nu _{iL}+\sum _{i=4,5,6}(K_K)_{la}
N_{aL}
\end{eqnarray}
The cross sections are characterized by the mass of the heavy lepton $M_N$
and the mixing parameters
\begin{eqnarray}
\ell\ell_{mix}&=&\sum_{a=4,5,6}(K_H)_{la}(K_H^\dagger)_{al}\;,
\end{eqnarray}
for $\ell = e,\mu,\tau$.
The existing constraints on the mixings are\cite{nrt}
\begin{eqnarray}
ee_{mix}\; \le && 0.0071\;, \\
\mu\mu_{mix}\; \le && 0.0014\;, \\
\tau\tau_{mix}\; \le && 0.033\;.
\end{eqnarray}
Since the best bound occurs for the second generation, the largest possible 
effects that are still allowed occur at an $e^+e^-$ collider.
The cross section depends on various mixing parameters\cite{kalyniakproc}:
\begin{eqnarray}
t_{mix}&=&|(K_L^*)_{li}(K_H)_{la}|^2\;, \\
s_{mix}&=&|(K_L^\dagger K_H)_{ia}|^2\;, \\
st_{mix}&=&(K_L^\dagger K_H)_{ia}(K_L)_{li}(K_H^*)_{la}\;,
\end{eqnarray}
which are bounded by the constraints listed above.

The single NHL production cross sections are displayed as a function of $M_N$
in Fig.~17 for $\sqrt{s} = 0.5, 1.0, 1.5$~TeV $e^+e^-$ colliders and for
$\sqrt{s} =0.5$~TeV $\mu^+\mu^-$ colliders. The maximal signal is smaller for
muon colliders because of the tighter constraint on $\mu\mu_{mix}$ relative
to $ee_{mix}$. In Fig.~18 the single NHL production cross sections are 
displayed for a $\sqrt{s} = 5$~TeV $e^+e^-$ collider and for a  
$\sqrt{s} = 4$~TeV $\mu^+\mu^-$ collider.

\begin{center}
\epsfxsize=1.0in
\hspace*{-1.2in}
\epsffile{snowmass.eps}

\vspace*{0.4in}
\parbox{5.5in}{\small  Figure 16: Feynman diagrams for 
$e^+e^-\to \overline{N_a}\nu _i$ in the (a) $s$-channel and 
the (b) $t$-channel (from Ref.~\cite{kalyniakproc}).}
\end{center}

\begin{center}
\vspace{-0.8in}
\epsfxsize=3.0in
\hspace*{0.9in}
\epsffile{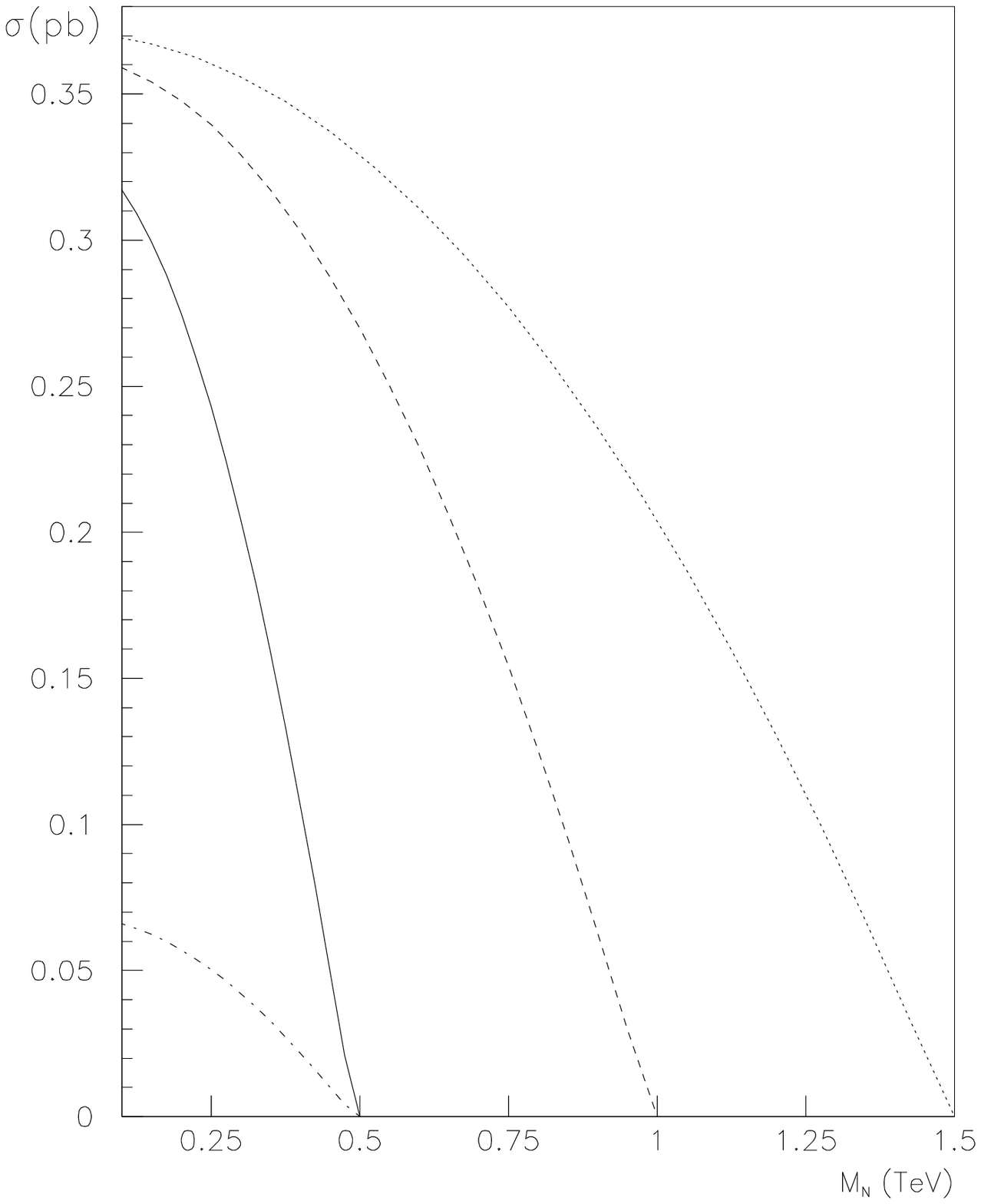}

\vspace*{0.7in}
\parbox{5.5in}{\small  Figure 17: Total cross section 
versus NHL mass $M_N$ for an
$e^+e^-$ collider at three different energies: $\sqrt{s}=0.5$~TeV (solid line),
$\sqrt{s}=1.0$~TeV (dashed line) and $\sqrt{s}=1.5$~TeV (dotted line), and for
a $\mu^+\mu^-$ collider at $\sqrt{s}=0.5$~TeV (dash-dotted line) 
(from Ref.~\cite{kalyniakproc}).}
\end{center}

\begin{center}
\vspace{-1.4in}
\epsfxsize=3.0in
\hspace*{0.9in}
\epsffile{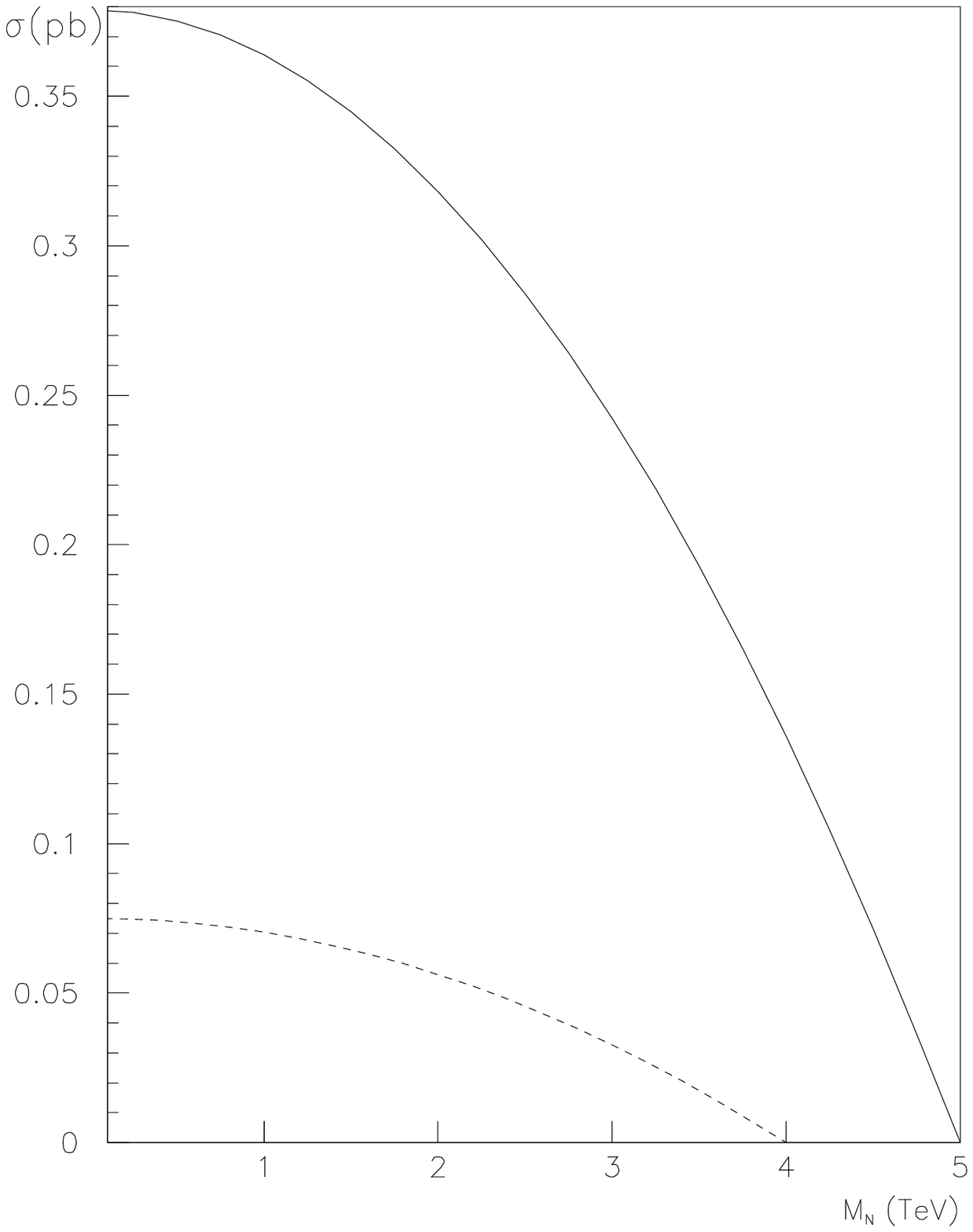}

\vspace*{0.4in}
\parbox{5.5in}{\small  Figure 18: Total cross section 
versus NHL mass $M_N$ for an
$e^+e^-$ collider at $\sqrt{s}=5.0$~TeV (solid line) , and for
a $\mu^+\mu^-$ collider at $\sqrt{s}=4.0$~TeV (dashed line) 
(from Ref.~\cite{kalyniakproc}).}
\end{center}

The discovery limits for NHL masses and mixings for the special case
$t_{mix}=s_{mix}=st_{mix}$ is shown in Table~V for the various machines
and integrated luminosities. The higher energy machines will be sensitive to 
mixings in the $10^{-5}$ to $10^{-6}$ range for much of the range of $M_N$.

\begin{table}[kalytab]
\centering
\begin{tabular}{lllr}
\hline
\hline
$\sqrt{s}({\rm TeV})$  & $L(fb^{-1})$  & $M_N(TeV)$ & $t_{mix}$ \\
\hline
1.0                    &  200  & 0.5   & $7 \times 10^{-6}$    \\
                       &       & 0.75  & $1 \times 10^{-5}$    \\
                       &       & 0.95  & $6 \times 10^{-5}$    \\
1.5                    &  200  & 0.5   & $5 \times 10^{-6}$    \\
                       &       & 1.0   & $9 \times 10^{-6}$    \\
                       &       & 1.25  & $2 \times 10^{-5}$    \\
                       &       & 1.45  & $8 \times 10^{-5}$    \\
4.0                    & 1000  & 0.5   & $9.5 \times 10^{-7}$  \\
                       &       & 1.0   & $1 \times 10^{-6}$    \\
                       &       & 2.0   & $1.2 \times 10^{-6}$  \\
5.0                    & 1000  & 0.5   & $9.4 \times 10^{-7}$  \\
                       &       & 1.0   & $9.5 \times 10^{-7}$  \\
                       &       & 2.0   & $1.1 \times 10^{-6}$  \\
\hline
\hline
\end{tabular}
\caption{Discovery limits for NHL masses and mixings 
(from Ref.~{\protect\cite{kalyniakproc}}).}
\label{kalytab}
\end{table}

\section{Conclusions}

The New Particles Subgroup concentrated on leptoquark signals at present and 
future colliders. The prospects for identifying the particle by measuring its 
properties was also addressed. Search strategies for bileptons at the NLC
and neutral heavy leptons at electron and muon colliders were described.

The signals for detection of these particles  falls into three classes:
direct detection by (1) single production or (2) pair production, or indirect
detection through their virtual effects. Whether or not any of these new
particles exists in nature is an open question, but the new colliders under
study at the Snowmass workshop will certainly extend the range far beyond the
existing limits.

\section*{Acknowledgement}

This work was supported in part by the U.S. Department of
Energy under Grant No. DE-FG02-95ER40661.

\end{document}